\documentclass[12pt,english,preprint,superscriptaddress]{revtex4-1}
\usepackage[latin9]{inputenc}
\usepackage{amsmath}
\usepackage{amssymb}
\usepackage{graphicx}
\usepackage{esint}
\usepackage{xcolor}
\usepackage{marginnote}
\usepackage{float}
\usepackage{color,soul}
\makeatletter
\usepackage{babel}
\usepackage{color}

\usepackage{ulem}
\usepackage{hyperref}

\newcommand{\strikeout}[1]{\ifmmode\text{\sout{\ensuremath{#1}}}\else\sout{#1}\fi}

\usepackage{babel}

\makeatother
\usepackage{babel}
\begin{document}
	
	\title{Diffusion caused by two noises - active and thermal}	
	\author{Koushik Goswami }
	\affiliation{Department of Inorganic and Physical Chemistry, Indian Institute of Science, Bangalore 560012, India}
	\author{K L Sebastian}
	\affiliation{Department of Inorganic and Physical Chemistry, Indian Institute of Science, Bangalore 560012, India}
	\affiliation{Indian Institute of Technology Palakkad, Ahalia Integrated Campus,	Kozhippara 678557, Palakkad, Kerala, India}
	\begin{abstract}
	The diffusion of colloids inside an active system - $e.g.$, within a living cell  or the dynamics of active particles itself ($e.g.$, self-propelled particles) can be modeled through overdamped Langevin equation which contains an additional noise term apart from the usual white Gaussian noise, originating from the thermal environment. The second noise is referred to as  \textquotedblleft active noise\textquotedblright as it arises from activity such as chemical reactions. The probability distribution function (PDF or the propagator) in space-time along with moments provides essential information for understanding their dynamical behavior. Here we employ the phase-space path integral method to obtain the  propagator, thereby moments and PDF for some possible models for such noise. At first, we discuss the diffusion of a free particle driven by active noise.  We consider four different possible models for active noise,  to capture the possible traits of such systems. We show that the PDF for systems driven by noises other than Gaussian noise largely deviates from normal distribution at short to intermediate time scales as a manifestation of out-of-equilibrium state, albeit converges to Gaussian distribution after a long time as a consequence of the central limit theorem. We extend our work to  the case of a  particle trapped in a harmonic potential and show that the system attains steady state at long time limit. Also, at short time scales, the nature of  distribution is different for different noises, $e.g.$, for particle driven by dichotomous noise, the probability is mostly concentrated near the boundaries whereas a long exponential tail is observed for a particle driven by  Poissonian white noise. 
	\end{abstract}
	\maketitle
	\tableofcontents
	\section{Introduction}
	
The dynamics of a colloidal particle in an environment of fast-moving solvent molecules is well described by the theory of	Brownian motion formulated by Einstein in $1905$ \cite{van1992stochastic,einstein1956investigations}.The motion of  the particle is random as it results from a large number of erratic collisions with the much smaller solvent molecules. The random force exerted by these molecules is uncorrelated at the timescale of relaxation of the particle and hence one usually models it as white, Gaussian noise (or thermal noise). Two interesting features
 of the resulting motion are: $(a)$ the mean square displacement of the particle is proportional to the time spent by the particle in the environment and $(b)$ the distribution of the particle's displacement is Gaussian (this follows from the central limit theorem) \cite{frey2005brownian}. In the recent past, situations where other types of noises, usually known as non-thermal noise, too act on the particle  have attracted
	quite a bit of attention. For example, in a living cell the cytosol
	is mainly composed of interlinked network system ($e.g.$ actin filaments)
	and motor proteins ($e.g.$ myosin II). With the consumption of ATP, the
	motors generate relative motion between the filaments resulting in
	random thrusts upon a particle residing inside the cell matrix. Such
	non-thermal noise, often termed as active noise, drives the system out of equilibrium. Contrary to thermal noise, such noise does not
	obey the fluctuation-dissipation theorem \cite{mizuno2007nonequilibrium}.
	Recently, several experiments \cite{toyota2011non,stuhrmann2012nonequilibrium,fodor2015activity}
	have studied the dynamics of a tracer particle ($in\,vitro$ as well
	as $in\,vivo$) in the cellular environment. In such experiments the
	fluctuations in spatial distribution of a probe particle placed inside
	the  ``active gel\textquotedblright  was studied using video microrheology.
	For all the cases studied, the spatial distribution largely deviates
	from the normal distribution.  Many a times, an exponential tail with a central Gaussian part \cite{toyota2011non,PhysRevLett.103.198103} has been observed.  Such behavior can be ascribed to a process driven by Poisson noise and thereby cytoplasmic matrix can be regarded as Poissonian bath.   Other than the noises
	which characterize the heat bath, a noise can be generated externally
	using some specific device. Such noise was introduced
	in the system in order to study non-equilibrium processes by Mestres
	\textit{et al.} \cite{mestres2014realization}. The noise had finite
	correlation time and hence it can be modeled as colored Gaussian noise
	(Ornstein-Uhlenbeck process) or as dichotomous noise. It may also be noted
	that dichotomous noise has a wide range of applicability, and has been used to model various physical processes such as stochastic resonance \cite{PhysRevE.60.1494,PhysRevE.62.R3031,PhysRevE.74.051115},
	synchronization effects \cite{doi:10.1021/jp004317x,doi:10.1143/PTP.116.819}, patterning \cite{PhysRevE.68.011103,PhysRevE.87.062924}, thermally activated transition between two states etc \cite{moss1989noise}.    Zheng \textit{et. al}  \cite{PhysRevE.88.032304}, studied the motion of micron-sized spherical Pt-silica Janus particles in H$_2$O$_2$ solution. They found that the probability distribution function for the displacement, deviates from Gaussianity under the experimental conditions, and the distribution is either a broadened peak or two-peaked structure depending upon the concentration of H$_2$O$_2$. They formulated a theoretical model by considering a Langevin equation with coupled translational and rotational motion. However one can circumvent the rotational degrees of freedom by replacing it with dichotomous noise.  Recently Malakar \textit{et al.}  have described  a one dimensional version of  the bacterial `Run and Tumble motion'(RTP) through a overdamped Langevin dynamics by adding an extra noise term \cite{Malakar2018}. In simplified 1D model, a bacteria can have either a state with $+u$ velocity or with $-u$, and the flip between these states is assumed to happen according to Poisson statistics. Not surprisingly, the extra noise has been considered as dichotomous.   Apart from bacterial motion, two state models have been  invoked extensively  in the context of gene expression   \cite{KEPLER20013116,PhysRevE.72.051907,shahrezaei2008analytical,Zeiser2009,PhysRevLett.113.268105,Thomas6994,PhysRevE.91.042111,potoyan2015dichotomous}. 
	  In particular, see Ref. \cite{potoyan2015dichotomous} in which the impact of gene switching on production of mRNA during transcription has been studied.  Depending on the binding or  unbinding of promoters  to  specific DNA sites, gene can  switch stochastically between two states, namely, active and inactive, and  the randomness arising due to such transitions has been modeled  as dichotomous noise.  Akin to  this, switching between discrete environmental states is pertinent in the study of predator-prey model \cite{LUO200769,luo2009stochastic,ZHU2009154}, evolution theory \cite{Ashcroft20140663,PhysRevLett.119.158301},  bacterial population growth \cite{kussell2005phenotypic,kussell2005bacterial} etc. Now coming to the non-equilibrium fluctuation in  diffusion process, there have been  a large number of experiments as well as theoretical investigations  on confined particles. For example, in Ref. \cite{PhysRevE.94.062150}, the dynamics of  an optically trapped Brownian particle coupled, to a bacterial bath has been studied to find that it exhibits non-Boltzmann (more specifically, heavy-tailed) stationary distribution in strong confinement regime, contrary to the force-free case.  However, for weak trapping and low concentration of bacteria the distribution becomes Gaussian  with an enhanced variance and thus in this regime,   force generated due to self-propulsion of bacteria acting on the passive particle can be taken to be correlated  Gaussian  noise \cite{PhysRevLett.113.238303}.  Another example is the run-and-tumble particles confined in a restricted region, which happen to accumulate near boundaries of the  geometry \cite{PhysRevE.90.062711,0295-5075-86-6-60002,PhysRevE.91.012125}. Driven systems have found  applications in other fields too, $e.g$, in hydrology to model soil water balance \cite{rodriguez1999probabilistic}, in ecology to describe the occurrence of fires in ecosystem \cite{d2006probabilistic}  etc.   Therefore it is of a great interest to calculate the time evolution
	of the probability distribution function (PDF), for a system that is subject to thermal noise (or immersed in a thermal bath), and is additionally driven by non-thermal
	noise. The common method for this purpose is to set up and solve the
	Fokker-Planck equation (FPE) describing the process \cite{Malakar2018}.   There have been some approximate as well as  exact solutions of the Fokker-Planck equations corresponding to the processes driven by  Poisson white noise, L\'evy stable noise and colored noise \cite{denisov2009generalized,haunggi1994colored}. However, the dynamical description for a system driven by two noises has not been explored in the full space-time range. Moreover, it is not always easy, particularly for colored noise, to solve the FPE and get the PDF analytically. Hence
	in this paper, we devise a general scheme  to cover a large set of noises by employing the phase-space path integral approach \cite{kamenev2011field,janakiraman2012path}. Further we investigate 
several dynamical systems corresponding to a wide class of physical processes, by  assuming different noise models, and obtain their dynamical features at transient and steady states.  Finally, we summarize our findings in the Section \ref{conclusion}. 
	
	\section{The phase-space path integral for a particle subject to two noises }
	
	A particle undergoing usual Brownian motion in a potential $U(x)$
	is described by the equation 
	\begin{equation}
	\zeta\frac{dx(t)}{dt}=-U'(x)+f(t),\label{eq:BrownianEquation}
	\end{equation}
	where the noise $f(t)$ is assumed to be Gaussian white noise having
	$<f(t)>=0$ and $<f(t)f(t')>=2k_{B}T\zeta\delta(t-t')$. The best
	way to characterize the noise $f(t)$ is through its probability
	density functional $\Psi[f(t)]\sim e^{-\frac{1}{4\zeta k_{b}T}\int dtf(t)^{2}}.$
	The diffusion coefficient of the particle $D$ is related to the coefficient
	of friction by $D=k_{B}T/\zeta$. It is convenient to rewrite the
	above as 
	\begin{equation}
	\frac{dx(t)}{dt}=-V'(x)+\eta(t),\label{eq:BrownianRewritten}
	\end{equation}
	with $V(x)=U(x)/\zeta$ and $\eta(t)=f(t)/\zeta$ so that $<\eta(t)\eta(t')>_{\eta(t)}=2D\delta(t-t')$.
	The noise $\eta(t)$ has the characteristic functional \cite{kamenev2011field}
	\begin{equation}
	\left<e^{i\int_{0}^{T}p(t)\eta(t)dt}\right>_{\eta(t)}=e^{-D\int_{0}^{T}p(t)^{2}dt},\label{eq:characteristic func.}
	\end{equation}
	where $<...>_{\eta(t)}$ denotes the average over all possible realizations
	of the noise $\eta(t)$.
	
	Now consider a situation where the particle is subjected to an additional
	noise $\sigma(t)$, due to the presence of active particles in the
	system or some other physical process. This modifies Eq. (\ref{eq:BrownianRewritten})
	to 
	\begin{equation}
	\frac{dx(t)}{dt}=-V'(x)+\eta(t)+\sigma(t).\label{eq:dynamics}
	\end{equation}
	Using the characteristic functional of Eq. (\ref{eq:characteristic func.}),
	one can express the probability density functional $P[\eta(t)]$
	as its functional Fourier transform: 
	\begin{equation}
	P[\eta]=\int Dp\,e^{-D\int_{0}^{T}p(t)^{2}dt}e^{-i\int_{0}^{T}p(t)\eta(t)dt}.\label{P(noise)}
	\end{equation}
	Using Eq. (\ref{eq:dynamics}) one can rewrite $P[\eta(t)]$ as a
	functional of $x(t)$ (for more details the reader is referred to
	\cite{janakiraman2012path}), to get the probability density functional
	for any path $x(t)$ of the particle, in the interval $t\in[0,T]$.
	\begin{equation}
	P[x]=\int Dp\,e^{-D\int_{0}^{T}p(t)^{2}dt}e^{-i\int_{0}^{T}p(t)(\overset{.}{x}(t)+V'(x))dt}\,e^{i\int_{0}^{T}p(t)\sigma(t)dt}.
	\end{equation}
	Note that as in \cite{janakiraman2012path,kamenev2011field}, we assume
	the Ito discretization of Eq. (\ref{eq:dynamics}) as a result of
	which the Jacobian of the transformation $\eta(t)\rightarrow x(t)$
	is just a constant. As the particle is not only driven by $\eta(t)$
	but also by $\sigma(t)$, one has to consider all the realizations
	of $\sigma(t)$ over the time period of $T$. Performing average over
	these, leads to 
	\begin{equation}
	P[x]=\int Dp\,e^{-D\int_{0}^{T}p(t)^{2}dt}e^{-i\int_{0}^{T}p(t)(\overset{.}{x}(t)+V'(x))dt}\,\left<e^{i\int_{0}^{T}p(t)\sigma(t)dt}\right>_{\sigma(t)}.
	\end{equation}
	For simplicity of notation, we have used $P[x]$ itself to denote
	the averaged quantity. Therefore, the probability of finding the particle
	at position $x_{f}$ after time $T$ provided it started at the position
	$x_{0}$ at the time $t=0$ is given by
	
	\begin{align}
	\mathbb{P}(x_{f},T;\,x_{0},0)  =  \int_{x(0)=x0}^{x(T)=x_{f}}Dx\int Dp&\,e^{-D\int_{0}^{T}p(t)^{2}dt}e^{-ip_{T}x_{f}+ip_{0}x_{0}}e^{i\int_{0}^{T}(x(t)\dot{p}(t)-V'(x)p(t))dt}\nonumber \\
	 &\times\left<e^{i\int_{0}^{T}p(t)\sigma(t)dt}\right>_{\sigma(t)}.\label{eq:finP1}
	\end{align}
	In Eq. (\ref{eq:finP1}), the functional integration is to be performed
	over all paths that start at $x_{0}$ at the time $t=0$ and end at
	$x_{f}$ at the time $t=T$. In cases where the potential $V(x)$
	is at the most quadratic, the path integration over $x$ can be performed
	easily (see \cite{janakiraman2012path} for more details) resulting
	in a Dirac delta functional involving $p(t)$. The delta functional
	makes the path integration over $p(t)$ easy to perform. The final
	result can be expressed as single integration over the value of $p$
	at the time $T$, $p_{T}\;(=p(T))$. We note that the same technique
	can be applied in underdamped limit and in that case path integration
	over $p(t)$ becomes a double integration over $p_{T}$ and $p_{0}\;(=p(0))$.
	As we will be using this procedure in the paper, we illustrate it
	for a free particle subject to the two noises. 
	
	\subsection{The free particle}
	
	Putting $V(x)=0$ in Eq. (\ref{eq:finP1}) we get the propagator for free particle as
	\begin{equation}
	\mathbb{P}(x_{f},T;\,x_{0},0)=\int_{x(0)=x0}^{x(T)=x_{f}}Dx\int Dp\,e^{-D\int_{0}^{T}p(t)^{2}dt}e^{-ip_{T}x_{f}+ip_{0}x_{0}}e^{i\int_{0}^{T}\,x(t)\dot{p}(t)dt}\mathcal{C}[p(t)]\label{eq:finP2}
	\end{equation}
	where $\mathcal{C}[p(t)]$ is the characteristic functional for the
	noise $\sigma(t)$ defined by 
	\begin{equation}
	\mathcal{C}[p(t)]=\left<e^{i\int_{0}^{T}p(t)\sigma(t)dt}\right>_{\sigma(t)}.\label{characterstic-functional-sigma}
	\end{equation}
	As $\int_{x(0)=x0}^{x(T)=x_{f}}Dx\,e^{i\int_{0}^{T}\,x(t)\dot{p}(t)dt}=\delta[\dot{p}(t)],$
	where $\delta[\cdots]$ is the Dirac delta functional, we get $\dot{p}(t)=0$
	and hence $p(t)=p_{T}=p_{0}$ . Thus the path integral over $p(t)$
	in Eq. (\ref{eq:finP2}) reduces to a single integration with respect
	to $p_{T}$. The propagator for the free particle becomes (after putting
	in appropriate normalization factor)
	\begin{equation}
	\mathbb{P}(x_{f},T;\,x_{0},0)=\frac{1}{2\pi}\int_{-\infty}^{+\infty}dp_{T}\,e^{-D\,p_{T}^{2}\,T}e^{-ip_{T}x_{f}+ip_{T}x_{0}}C(p_{T}),\label{eq:finP3}
	\end{equation}
	where the function $C(p_{T})$ is defined by $C(p_{T})=\mathcal{C}[p(t)]_{p(t)\equiv p_{T}}$. 
	
	\subsection{Particle in a harmonic trap}
	
	In this subsection, we analyze the dynamics of a particle bound in
	a harmonic potential $V(x)=\frac{1}{2}\lambda x^{2}$, where $\lambda$
	is the stiffness constant. In this case, the path integration over
	$x$ in Eq. (\ref{eq:finP1}) can be performed and results in the
	delta functional $\delta[\dot{p}(t)-\lambda\,p(t)]$. This implies
	$\dot{p}(t)=\lambda p(t)$ which on solution gives $p(t)=p_{T}e^{\lambda(t-T)}$
	where $p_{T}=p(T)$. Therefore the propagator for a harmonic oscillator
	becomes
	\begin{equation}
	\mathbb{P}(x_{f},T;\,x_{0},0)=\frac{1}{2\pi}\int_{-\infty}^{+\infty}dp_{T}\,e^{-\frac{D\,p_{T}^{2}}{2\lambda}(1-e^{-2\lambda T})-ip_{T}x_{f}+ip_{0}x_{0}}\,G(p_{T}).\label{eq:finP3sho}
	\end{equation}
	Here 
	\begin{equation}
	G(p_{T})=\mathcal{C}[p(t)]_{p(t)\equiv p_{T}\,e^{-\lambda(T-t)}}=\left<e^{ip_{T}\int_{0}^{T}\,e^{-\lambda(T-t)}\sigma(t)dt}\right>_{\sigma(t)}.\label{Gsho}
	\end{equation}
	Now if the initial distribution of the particle is $P(x_{0})$, then
	one can get the probability distribution function for finding the
	particle at position $x_{f}$ after time $T$, by integrating over
	all possible initial positions $x_{0}$. Thus the probability distribution
	becomes 
	\begin{equation}
	\mathbb{P}(x_{f},T)=\int_{-\infty}^{+\infty}dx_{0}\,\mathbb{P}(x_{f},T;\,x_{0},0)\,P(x_{0}).\label{Probfinal}
	\end{equation}
	We now illustrate the method for several different choices of $\sigma(t).$
	
	\section{Free particle subject to different types of noises $\sigma(t)$}
	
	\subsection{ Gaussian colored Noise }
	
	The simplest case that one can have is: $\sigma(t)$ is  Gaussian colored
	noise $\sigma_{CG}(t)$ with zero mean. The correlation function is
	then given by $<\sigma_{CG}(t)\sigma_{CG}(t')>=\frac{D_{A}}{\tau}\,e^{-\frac{\vert t-t'\vert}{\tau}}.$
	Using the cumulant expansion technique \cite{van1992stochastic} one
	can find its characteristic functional to be
    \begin{equation}
	\left<e^{i\int_{0}^{T}p(t)\sigma_{CG}(t)dt}\right>_{\sigma_{CG}(t)}=e^{-\frac{1}{2}\int_{0}^{T}\int_{0}^{T}p(t)p(t')<\sigma_{CG}(t)\sigma_{CG}(t')>dt'\,dt}.\label{ch gaussian}
	\end{equation}
	This gives 
	\begin{equation}
	C(p_{T})=e^{-p_{T}^{2}D_{A}[T-\tau(1-e^{-\frac{T}{\tau}})]}.\label{ch gaus}
	\end{equation}
	Using Eq. (\ref{ch gaus}) in Eq. (\ref{eq:finP3}), one can get the
	probability distribution function 
	\begin{equation}
	\mathbb{P}(x_{f},T;\,x_{0},0)=\frac{e^{-\frac{(x_{f}-x_{0})^{2}}{4T[D+D_{A}(1-\frac{\tau}{T}(1-e^{-\frac{T}{\tau}}))]}}}{\sqrt{4\pi T[D+D_{A}(1-\frac{\tau}{T}(1-e^{-\frac{T}{\tau}}))]}}.
	\end{equation}
	Not surprisingly, the probability distribution is Gaussian with a
	mean squared displacement $<x^{2}(T)>=2T\left[D+D_{A}(1-\frac{\tau}{T}(1-e^{-\frac{T}{\tau}}))\right]$.
	In the very small time limit (i.e., $T\ll\tau)$, $<x^{2}(T)>=2DT$
	showing that at such short time scales, the particle is moving with
	white noise-only diffusivity $D$. In the very large time limit, $<x^{2}(T)>=2(D+D_{A})T$,
	implying that the diffusion has an effective diffusivity $D+D_{A}$.
	As the noises acting upon the particle are Gaussian in nature, it
	is natural to have Gaussian distribution of particle's position over
	any timescale of observation.
	
	\subsection{Dichotomous Poisson Noise}
	
\begin{figure}[H]
\includegraphics{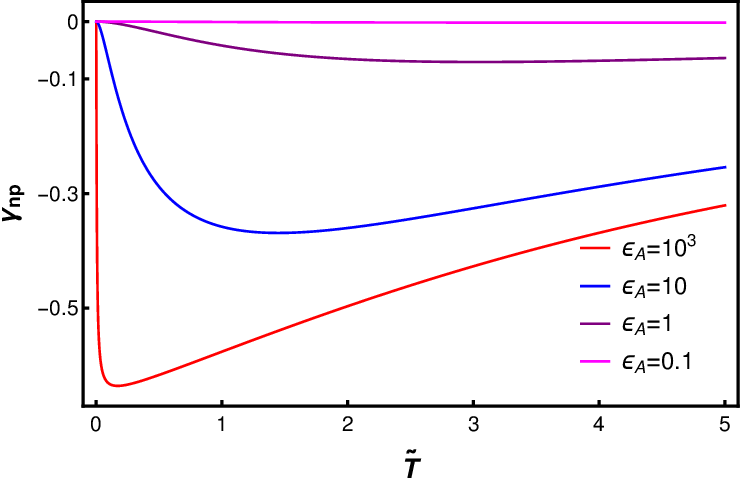} 
\caption{Dichotomous plus thermal noise: The non-Gaussian parameter ($\gamma_{np}$) is plotted against the dimensionless time $\tilde{T} $ for different diffusivity ratios $\epsilon_{A}=D_A/D$.  }
\label{ngpplotdicho} 
\end{figure}

	 As we have mentioned before, the motion of self-propelled particles (Janus microbead or bacteria) can be conceived through Eq. (\ref{eq:dynamics}), considering  $\sigma(t)$ as dichotomous Poissonian noise. We will denote such noise by $\sigma_{DP}(t)$.  It takes only two values: $+ u$
	and $- u$. Further, it is assumed to have the correlation $<\sigma_{DP}(t)\sigma_{DP}(t')>=u^{2}\,e^{-\gamma\mid t-t'\mid}$,
	implying the noise has a finite correlation time equal to $1/\gamma$.
	Note that $D_{A}=u^{2}/\gamma$ has the dimension of diffusivity,
	and can be thought of as diffusivity under the dichotomous noise.
	The characteristic function for such noise is given by \cite{PhysRevA.41.754,Klyatskin1977,doi:10.1142/S0217979206034881}
	
	\begin{equation}
	C(p_{T})=\left<e^{ip_{T}\int_{0}^{T}\sigma_{DP}(t)dt}\right>_{\sigma_{DP}(t)}=e^{-\frac{\gamma T}{2}}\left[\text{cosh}\left(\frac{\gamma_{u}T}{2}\right)+\frac{\gamma}{\gamma_{u}}\text{sinh}\left(\frac{\gamma_{u}T}{2}\right)\right],\label{eq:ch func dicho}
	\end{equation}
	where 
	\begin{align}
	\gamma_{u}  = & \gamma\left[1-\frac{4 u^{2}p_{T}^{2}}{\gamma^{2}}\right]^{\frac{1}{2}}\;\hspace{0.5cm}\mbox{ if }\;\frac{4 u^{2}p_{T}^{2}}{\gamma^{2}}\;\leq\;1,\nonumber \\
	 = & \gamma\left[\frac{4 u^{2}p_{T}^{2}}{\gamma^{2}}-1\right]^{\frac{1}{2}}\frac{p_{T}}{|p_{T}|}\;\hspace{0.5cm}\mbox{otherwise.}
	\end{align}
	Using Eq. (\ref{eq:ch func dicho}) in Eq. (\ref{eq:finP3}) one can
	get the PDF:
	\begin{equation}
	\mathbb{P}(x_{f},T;\,x_{0}=0,0)=\frac{1}{2\pi}\int_{-\infty}^{+\infty}dp_{T}\,e^{-ip_{T}x_{f}}\,F(p_{T}),\label{eq:dicho P}
	\end{equation}
	where 
	\begin{equation}
	F(p_{T})=e^{-DT\,p_{T}^{2}}e^{-\frac{\gamma T}{2}}\left[\text{cosh}\left(\frac{\gamma_{u}T}{2}\right)+\frac{\gamma}{\gamma_{u}}\text{sinh}\left(\frac{\gamma_{u}T}{2}\right)\right].\label{dicho F}
	\end{equation}
	Here we have taken the initial position $x_{0}=0$ without any loss
	of generality. The Fourier transform of Eq. (\ref{eq:dicho P}) can
	be written as a convolution. The $m^{th}$ moment of the displacement
	at any instant can be evaluated easily, using
	
	\begin{equation}
	<x^{m}(T)>=\left[(-i)^{m}\frac{d^{m}F(p_{T})}{dp_{T}^{m}}\right]_{p_{T}=0}.\label{m-th moment}
	\end{equation}
	Using Eq. (\ref{dicho F}) and Eq. (\ref{m-th moment}) we get
	
	\begin{equation}
	<x^{2}(T)>=2DT+\frac{2D_{A}}{\gamma}(-1+T\gamma+e^{-T\gamma}).
	\end{equation}
  We first consider the case $\gamma T\ll 1$.   Then $\left < x^2 (T)\right > \approx 2DT+\gamma D_A T^2=DT(2+u^2T/D)    $.  Two limits can now be distinguished.  The first is when $1\gg u^2T/2D$, in which case $\left < x^2 (T)\right > \approx 2DT$, meaning that the motion is purely diffusive.  The other is when $u^2T/2D\gg 1$, leading to $\left < x^2 (T)\right >\approx u^2T^2$, and the motion is ballistic in this limit. In the long time limit, $<x^{2}(T)>=2DT+2D_{A}T$ and the motion is purely diffusive with a diffusivity $D+D_A$.  Insight into the behavior of probability distribution can be gained from the fourth moment,
	through the non-Gaussian parameter (NGP) $\gamma_{np}$ defined by 
	\begin{equation}
	\gamma_{np}=\frac{<x^{4}(T)>}{3<x^{2}(T)>^{2}}-1.\label{NGP}
	\end{equation}
	For a Gaussian distribution $\gamma_{np}$ is zero whereas the distribution
	which decays faster than Gaussian has negative $\gamma_{np}$ value
	and slower decaying distribution has a positive $\gamma_{np}$. After
	calculating the moments from Eq. (\ref{eq:ch func dicho}) and Eq.
	(\ref{m-th moment}) and using them in Eq. (\ref{NGP}) we have 
	\begin{align}
	\gamma_{np}&=\frac{5-2T\gamma-4e^{-T\gamma}(1+T\gamma)-e^{-2T\gamma}}{[T\gamma+\frac{D}{D_{A}}T\gamma+e^{-T\gamma}-1]^{2}}\nonumber\\
	&=\frac{5-2\tilde{T}-4e^{-\tilde{T}}(1+\tilde{T})-e^{-2\tilde{T}}}{[\tilde{T}+\frac{\tilde{T}}{\epsilon_A}+e^{-\tilde{T}}-1]^{2}}
	\label{eq:NGP dicho}
	\end{align}
	We use the dimensionless variables, $\tilde{T}=\gamma T$, $\epsilon_A=\frac{D_A}{D}$ and have plotted $\gamma_{np}$ in Fig. \ref{ngpplotdicho}
	
 For any finite value of $\epsilon_A$, in both large and small time limits $\gamma_{np}(\tilde{T})$ approaches the value of zero indicating its Gaussian behavior.   However, for $\epsilon_A \rightarrow \infty\; (D\rightarrow 0)$, $\gamma_{np}(0)=-2/3$ as in this limit only the dichotomous noise is important. In Fig. \ref{ngpplotdicho}, the plot for $\epsilon_A=1000$ exemplifies this kind of behavior.   In this case the value of $\gamma_{np}(\tilde{T})$ increases with time, eventually attaining the value zero in the long time limit.
	
	$\gamma_{np}(\tilde{T})$ is always
	is negative.  This is essentially because the dichotomous noise makes
	the distribution non-Gaussian. Further, the fact that it is always negative means that the distribution is narrower than a Gaussian of the same $<x^2>$ value.  To access the
	full description of the dynamics one is required to obtain PDF which
	is spatial Fourier inverse of $F(p_{T})$. Obviously, 
	\begin{equation}
	\mathbb{P}(x_{f},T;\,0,0)=\int_{-\infty}^{+\infty}\,dx'\,F_{1}(x_{f}-x')F_{2}(x'),\label{G}
	\end{equation}
	where 
	\begin{equation}
	F_{1}(x)=\frac{1}{2\pi}\int_{-\infty}^{+\infty}dp_{T}\,e^{ip_{T}x}e^{-D\,T\,p_{T}^{2}}=\frac{1}{\sqrt{4\pi DT}}e^{-\frac{x^{2}}{4DT}}\label{g1}
	\end{equation}
	and 
	\begin{align}
	F_{2}(x) & =  \frac{1}{2\pi}\int_{-\infty}^{+\infty}\,dp_{T}\,e^{ip_{T}x}e^{-\frac{\gamma T}{2}}[\text{cosh}(\frac{\gamma_{u}T}{2})+\frac{\gamma}{\gamma_{u}}\text{sinh}(\frac{\gamma_{u}T}{2})]\nonumber \\
	& = \frac{e^{-\frac{\gamma T}{2}}}{2}\,[\delta(x+u T)+\delta(x-u T)]\nonumber \\
	&   +\frac{\gamma}{4 u}e^{-\frac{\gamma T}{2}}\left[I_{0}\left(\frac{\gamma T}{2}y\right)+\frac{1}{y}\,I_{1}\left(\frac{\gamma T}{2}y\right)\right][\theta(x+u  T)-\theta(x-u  T)]\label{g2},
	\end{align}
	where $I_{\nu}$ is the modified Bessel Function and $y=\sqrt{1-\left(\frac{x}{u T}\right)^{2}}$.
	
	Apart from the two dimensionless parameters $\tilde{T}$, $\epsilon_A$, we now introduce another dimensionless quantity  which is defined as: $\tilde{x}=x\frac{\gamma}{u}$  and thereby we rewrite Eq. (\ref{G})-(\ref{g2}) in terms of them as follows:
	\begin{equation}
	\mathbb{P}(\tilde{x}_{f},\tilde{T};\,0,0)=\int_{-\infty}^{+\infty}\,d\tilde{x'}\,F_{1}(\tilde{x}_{f}-\tilde{x'})F_{2}(\tilde{x'}),\label{G'}
	\end{equation}
	with
	\begin{equation}
	F_{1}(\tilde{x})=\sqrt{\frac{\epsilon_A}{4\pi \tilde{T}}}\,e^{-\frac{\epsilon_A\,\tilde{x}^{2}}{4 \tilde{T}}}\label{g1'}
	\end{equation}
	and
	\begin{align}
	F_{2}(\tilde{x}) &  = \frac{e^{-\frac{\tilde{ T}}{2}}}{2}\,[\delta(\tilde{x}+\tilde{ T})+\delta(\tilde{x}-\tilde{ T})]\nonumber \\
	&   +\frac{1}{4 }e^{-\frac{\tilde{ T}}{2}}\left[I_{0}\left(\frac{\tilde{ T}}{2}y\right)+\frac{1}{y}\,I_{1}\left(\frac{\tilde{ T}}{2}y\right)\right][\theta(\tilde{x}+\tilde{ T})-\theta(\tilde{x}- \tilde{T})]\label{g2'}
	\end{align}
	where
	$y=\sqrt{1-\left(\frac{\tilde{x}}{\tilde{ T}}\right)^{2}}$.   These results are identical to those of Malakar {\it et al.} \cite{Malakar2018}.
	
\begin{figure}	\includegraphics[width=0.5\linewidth]{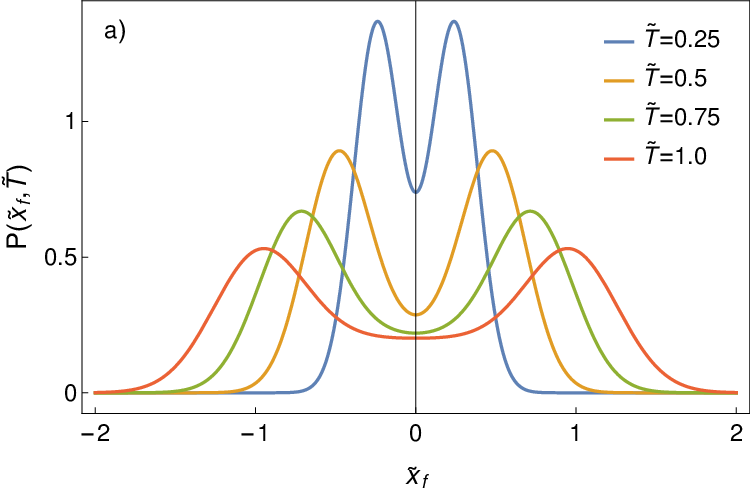} \quad{}
\includegraphics[width=0.5\linewidth]{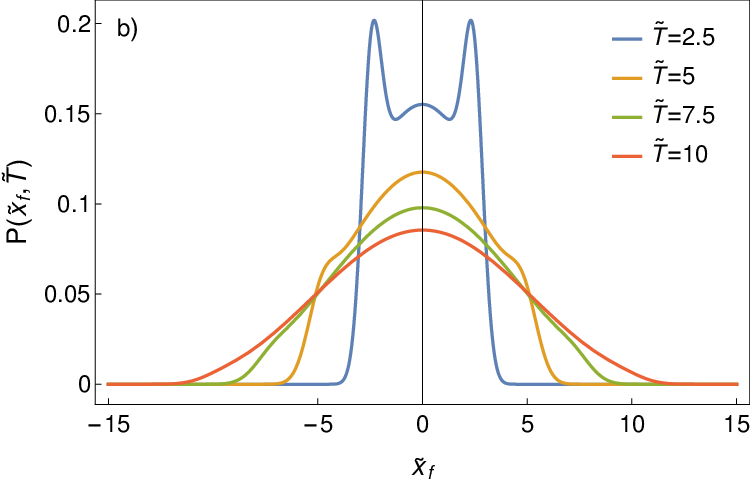} 
\caption{Dichotomous+thermal noise: The probability distribution function vs. displacement plot at (a) small $\tilde{T}$ and (b) large $\tilde{T}$ limit for high diffusivity ratio at $\epsilon_A=D_A/D=25$. }
\label{p(x) dicho plot highd} 
\end{figure}

\begin{figure}	\includegraphics[scale=1]{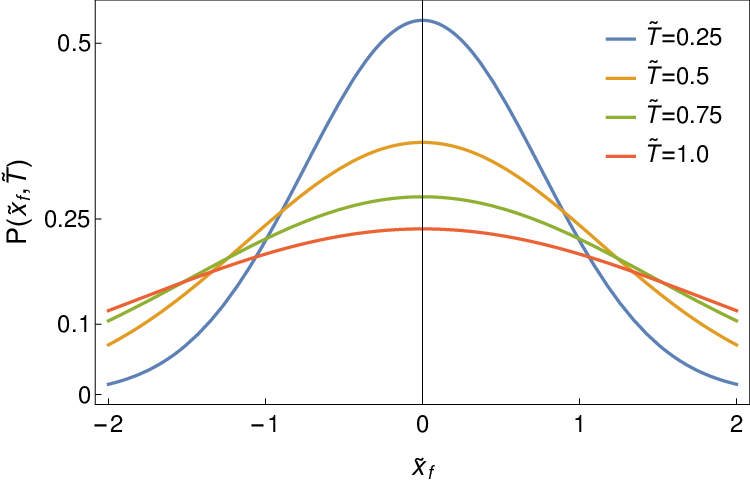}
\caption{Dichotomous+white noise: The probability distribution function is plotted against displacement at different times for small diffusivity ratio at $\epsilon_A=D_A/D=1$.}\label{p(x) dicho plot lowd} 
\end{figure}

	Further mathematical details are discussed in Appendix \ref{appen1}.
	From Eq. (\ref{P dicho small T}) one sees that during times lesser
	than the correlation time of dichotomous noise, the distribution function
	is centered around $\pm u T$ as shown in Fig. \ref{p(x) dicho plot highd} (a). This is due to  the fact that dichotomous noise itself
	can take up only two values $+ u$ and $- u$, resulting particles to accumulate at those regions at initial stage. However in the case where strength of active noise is smaller compared to white noise these two peaks can not be distinguished as shown in Fig. \ref{p(x) dicho plot lowd}. With the passage of time the distribution spreads as an effect of white noise as well as the
	active noise which causes an additional diffusivity $D_{A}$. After
	a very long time the distribution becomes Gaussian as given in Eq.
	(\ref{P dicho large T}). 
\subsection{Poissonian White Noise}
	
	\label{PWNsection}
	
	Here we consider $\sigma(t)$ as Poissonian white noise denoted by
	$\sigma_{PW}(t)$ - this means that the noise is generated by a random
	sequence of pulses at times $t_{i}$ that follow Poisson distribution.
	We can express it as $\sigma_{PW}(t)=\sum_{i}a_{i}\,g(t-t_{i})$,
	where $g(t-t_{i})$ is the pulse centered at $t_{i}$ which has an
	amplitude $a_{i}$. The probability of having $n$ pulses in a time
	interval $T$ is given by 
	\begin{equation}
	P(n;\mu,T)=\frac{(\mu T)^{n}e^{-\mu T}}{n!}.
	\end{equation}
	In the above, $\mu$ is the rate of the pulses. Usually $g(t-t_{i})$ is assumed to be a delta function pulse and the amplitude
	$a_{i}$ can take any value between $-\infty$ to $+\infty$. We assume
	it to be (double-)exponentially distributed (more specifically distributed
	according to the Laplace distribution). Hence the probability distribution
	of amplitude $a$, denoted as $\mathtt{P}(a)$ can be expressed as  
	\begin{equation}
	\mathtt{P}(a)=\frac{1}{2\,a_{0}}e^{-\frac{\mid a\mid}{a_{0}}},\label{eq:P(a)}
	\end{equation}
	where $a_{0}$ determines how broad the distribution of $a$ is. The Poisson
	white noise has zero mean and delta correlation. That is,
	\begin{align}
	<\sigma_{PW}(t_{1})\sigma_{PW}(t_{2})>=2\,\mu a_{0}^{2}\,\delta(t_{1}-t_{2}). \label{correlationpw}
	\end{align}
	The strength of correlation, $\mu a_{0}^{2}$ can be regarded as diffusivity
	due to the Poisson noise and hence will be denoted as $D_{A}$.
	
\begin{figure}[H]
\includegraphics{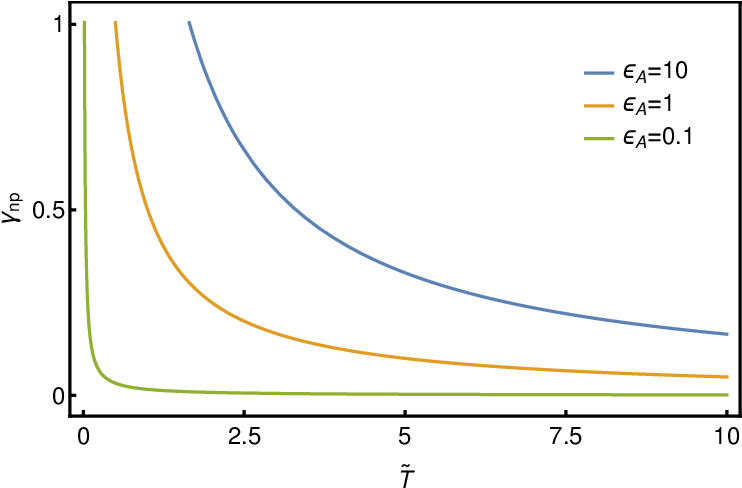} 
\caption{Gaussian+Poissonian white noises: the non-Gaussian parameter ($\gamma_{np}$) is plotted against the rescaled time $\tilde{T}$ for different values of $\epsilon_A=D_A/D$. }
\label{ngpplotpoisson} 
\end{figure}

	Such noise can be characterized  by its characteristic
	functional (see  Ref. \cite{goswami2019heat} for more information): 
	\begin{equation}
	\left<e^{i\int_{0}^{T}q(s)\,\sigma_{PW}(s)\,ds}\right>_{\sigma_{PW}(s)}=\text{exp}\left[-\mu\int_{0}^{T}\frac{a_{0}^{2}q(s)^{2}}{1+a_{0}^{2}q(s)^{2}}ds\right].\label{eq:ch functional Poisson}
	\end{equation}
	To get the PDF from the Eq. (\ref{eq:finP3}), one is required to know the
	characteristic function which is obtained in this case from Eq. (\ref{eq:ch functional Poisson})
	and it is 
	\begin{equation}
	C(p_{T})=\exp\left(-\frac{T}{\tau_{A}}\frac{a_{0}^{2}p_{T}^{2}}{1+a_{0}^{2}p_{T}^{2}}\right).\label{ch PSN}
	\end{equation}
	Here the characteristic time scale $\tau_{A}$ for active noise is
	the inverse of Poisson rate, $i.e.$, $\tau_{A}=\frac{1}{\mu}$. Therefore,
	the PDF can be expressed as 
	\begin{equation}
	\mathbb{P}(x_{f},T;\,x_{0}=0,0)=\frac{1}{2\pi}\int_{-\infty}^{+\infty}dp_{T}\,e^{-ip_{T}x_{f}}\,e^{-D\,T\,p_{T}^{2}-\frac{T}{\tau_{A}}\frac{a_{0}^{2}p_{T}^{2}}{1+a_{0}^{2}p_{T}^{2}}}.\label{eq:Poisson P}
	\end{equation}
	The MSD is calculated using the relation (\ref{m-th moment}) and is given
	by: $<x(T)^{2}>=2DT+2D_{A}T$. So the dynamics
	is always Fickian. The non-Gaussian parameter ($\gamma_{np}$) is computed
	from Eq. (\ref{NGP}) using Eq. (\ref{eq:Poisson P}) and it is given
	by 
	\begin{equation}
	\gamma_{np}=\frac{2a_{0}^{2}D_{A}}{T(D+D_{A})^{2}}=\frac{2\epsilon_{A}^{2}}{\tilde{T}(1+\epsilon_{A})^{2}}\label{ngppoissonexpn},
	\end{equation}
	where $\tilde{T}=T/\tau_{A}$ is the dimensionless time
	and $\epsilon_{A}=D_{A}/D$. Using Eq. (\ref{ngppoissonexpn}) we
	have made a plot of non-Gaussian parameter $(\gamma_{np})$ as a function
	of scaled time $\tilde{T}$ as shown in Fig. \ref{ngpplotpoisson}. 
	One can observe that at very large time limit $\gamma_{np}$ decays
	to zero evincing its convergence towards the Gaussian distribution.
	However in the very small time limit the distribution is not Gaussian
	as $\gamma_{np}$ is non-zero. But it decays to zero very rapidly
	for the case where normal diffusivity is very high in magnitude compared
	to active diffusivity.  It is interesting to note that in comparison with the dichotomous noise considered in the previous section, for this process $\gamma_{np}(t)$ is always positive.  This means that the distribution is always broader than a Gaussian distribution and eventually reaches Gaussian in the long time limit.  This happens only slowly, as $\gamma_{np} \; \propto\; 1/\tilde{T}$.
	We will now find the short time and long time limits of the PDF. Considerable
	insight into the nature of PDF is gained from rewriting the probability
	distribution as in Eq. (\ref{rewriting}). For this, we scale the variables into  dimensionless quantities as:  $\tilde{x}_f=\frac{x_f}{a_0}$, $\tilde{p_T}=a_0\,p_T$  and write the
	Eq. (\ref{eq:Poisson P}) as
	\begin{align}
	\mathbb{P}(\tilde{x}_{f},\tilde{T};\,\tilde{x}_{0}  =0,0)=\frac{1}{2\pi}\int_{-\infty}^{+\infty}d\tilde{p}_{T}\,e^{-i\tilde{p}_{T}\tilde{x}_{f}}\,e^{-\frac{D}{D_A}\,\frac{T}{\tau_A}\,\tilde{p}_{T}^{2}-\frac{T}{\tau_{A}}(1-\frac{1}{1+\tilde{p}_{T}^{2}})} \\
	 =\frac{1}{2\pi}\int_{-\infty}^{+\infty}d\widetilde{p}_{T}\,e^{-i\widetilde{p}_{T}\;\widetilde{x}_{f}}\,e^{-\,\frac{\widetilde{T}\,}{\epsilon_{A}}\widetilde{p}_{T}^{2}-\tilde{T}}\sum_{n=0}^{\infty}\frac{1}{n!}\left(\frac{\widetilde{T}}{1+\widetilde{p}_{T}^{2}}\right)^{n}\label{rewriting_0}\\
	 =\frac{1}{2\pi}\int_{-\infty}^{+\infty}d\widetilde{p}_{T}\,e^{-i\widetilde{p}_{T}\tilde{x}_{f}}\,e^{-\frac{\widetilde{T}\,}{\epsilon_{A}}\widetilde{p}_{T}^{2}-\widetilde{T}}+Q\nonumber \\
	 =\frac{\sqrt{\epsilon_{A}}}{\sqrt{4\pi\tilde{T}}}e^{-\tilde{T}-\widetilde{x}_{f}^{2}\epsilon_{A}/(4\tilde{T})}+Q.\label{rewriting}
	\end{align}
	
	In the above, 
	\begin{equation}
	Q=\frac{1}{2\pi}\int_{-\infty}^{+\infty}d\widetilde{p}_{T}\,e^{-i\widetilde{p}_{T}\;\widetilde{x}_{f}}\,e^{-\,\frac{\widetilde{T}\,}{\epsilon_{A}}\widetilde{p}_{T}^{2}-\tilde{T}}\sum_{n=1}^{\infty}\frac{1}{n!}\left(\frac{\widetilde{T}}{1+\widetilde{p}_{T}^{2}}\right)^{n}.\label{Qpwn}
	\end{equation}
	{One can use the  convolution theorem  (Eq. (\ref{G})) as the following
		Fourier transforms are well known: 
		\begin{equation}
		\int_{-\infty}^{+\infty}d\widetilde{p}_{T}\,\frac{e^{i\widetilde{p}_{T}x_{f}}}{\left(1+\widetilde{p}_{T}^{2}\right)^{\nu}}=\frac{2^{\frac{3}{2}-\nu}\,\sqrt{\pi}\,\left|\widetilde{x}_{f}\right|^{\nu-\frac{1}{2}}K_{\nu-\frac{1}{2}}\left(\left|\widetilde{x}_{f}\right|\right)}{\Gamma(\nu)}\label{fouriertransformCauchy}
		\end{equation}
		and 
		\begin{equation}
		\int_{-\infty}^{+\infty}d\widetilde{p}_{T}\,e^{i\widetilde{p}_{T}(\widetilde{x}_{f}-x)-\frac{\widetilde{T}\,}{\epsilon_{A}}\widetilde{p}_{T}^{2}}=\sqrt{\frac{\epsilon_{A}}{4\pi\tilde{T}}}\,e^{-\frac{\epsilon_{A}(\widetilde{x}_{f}-x)^{2}}{4\tilde{T}}}\label{fouriertransformGaussian},
		\end{equation}
		where $K(x)$ is Modified Bessel function of second kind \cite{bateman1953higher}. 
		Then one gets
		\begin{align}
		\mathbb{P}&(\tilde{x}_{f},\tilde{T};\,\tilde{x}_{0}  =0,0)=\sqrt{\frac{\epsilon_{A}}{4\pi\tilde{T}}}e^{-\tilde{T}-\epsilon_{A}\tilde{x_{f}^{2}}/(4\tilde{T})}\nonumber \\
		& +\frac{1}{2\pi}\sum_{n=1}^{\infty}\frac{e^{-\tilde{T}}}{n!}\left(\tilde{T}\right)^{n}\,\int_{-\infty}^{\infty}dx\frac{2^{\frac{3}{2}-n}\,\sqrt{\pi}\,\left|\tilde{x}\right|^{n-\frac{1}{2}}K_{n-\frac{1}{2}}\left(|\tilde{x}|\right)}{\Gamma(n)}\sqrt{\frac{\epsilon_{A}}{4\pi\tilde{T}}}e^{-\epsilon_{A}\tilde{(x_{f}}-x)^{2}/(4\tilde{T})}\label{p(x)pwnsmallx2},
		\end{align}
		which is an exact expression for the probability distribution function. We use this to demonstrate that the PDF would have an exponential
		(and not Gaussian) tail in the appropriate limits. 
		\subsubsection{Short time limit}
		We now consider the limit where active noise is the dominant effect
		- i.e., the displacement due to the thermal noise is much smaller than the average
		jump due to active noise, i.e., $\epsilon_{A}/\tilde{T}\gg1$. Then, one can
		evaluate the convolution integral by approximating $\sqrt{\frac{\epsilon_{A}}{4\pi\tilde{T}}}e^{-\epsilon_{A}\tilde{(x_{f}}-x)^{2}/(4\tilde{T})}\approx\delta(\tilde{x}_{f}-x)$.
		In the large $\tilde{x}{}_{f}$ limit, (i.e., $\tilde{x}_{f}\gg1$),
		one can use the asymptotic form for the Bessel function $K_{n}\left(|\tilde{x}_{f}|\right)\approx\sqrt{\frac{\pi}{2\mid \tilde{x}_{f}\mid}}\,e^{-\mid \tilde{x}_{f}\mid}$.
		With this, the sum on the right-hand side of Eq. (\ref{p(x)pwnsmallx2})
		can be performed to get 
		\begin{equation}
		\mathbb{P}(\tilde{x}_{f},\tilde{T};\,\tilde{x}_{0}=0,0)=\sqrt{\frac{\epsilon_{A}}{4\pi\tilde{T}}}e^{-\tilde{T}-\epsilon_{A}\tilde{x_{f}^{2}}/(4\tilde{T})}+\frac{e^{-|\tilde{x}_{f}|-\tilde{T}}}{2}\sqrt{\frac{2\,\tilde{T}}{\mid\tilde{x}_{f}\mid}}I_{1}\left(\sqrt{2\tilde{T}\mid\tilde{x_{f}}\mid}\right)\label{p(x)pwnsmallfin},
		\end{equation} 
		where $I_{1}$ is the modified Bessel function of the first kind \cite{bateman1953higher}. In the
		limit that we have just considered, it is clear that the first term on
		the RHS can be neglected. For times such that $2\tilde{T}\mid\tilde{x_{f}}\mid\gg1$,
		one can use the asymptotic expansion of $I_{1}(z)\approx\frac{e^{z}}{\sqrt{2\pi z}}$
		to get 
		\begin{equation}
		\mathbb{P}(\tilde{x}_{f},\tilde{T};\,\tilde{x}_{0}=0,0)\approx\,\frac{\tilde{T}}{2\pi}\,\frac{e^{-\vert \tilde{x}_{f} \vert +\sqrt{2\tilde{T}|\tilde{x}_{f}|}-\tilde{T}}}{\left(2\tilde{T}\,|\tilde{x}_{f}|\right)^{\frac{3}{4}}}. \label{asymptoticp}
		\end{equation}
		Therefore one can state that the distribution given in Eq. (\ref{p(x)pwnsmallfin})
		consists of two terms - the first term manifests the Gaussian distribution
		whereas the second term does for exponential one. For the system with
		$D\ll D_{A}$, at very short time period it attains the condition: $x_{f}^{2}\gg DT$
		at very small displacement limit. Hence the distribution exhibits
		a prominent, long exponential tail. But for the system with $D\gg D_{A}$,
		the same condition is achieved at long displacement limit resulting in
		an  insignificant exponential tail, which disappears quickly as time
		progresses. On the other hand Gaussian is dominant feature here as
		it decays very slowly. This result supports our previous analysis
		on non-Gaussian parameter.
		
	We have performed numerical integration of Eq. (\ref{rewriting}) for two different values of $\epsilon_{A}$ and have made	a plot of PDF  as a function of dimensionless displacement $\tilde{x}_f$ within a short time range $\tilde{T}$ as shown in Fig. \ref{p(x)Poisson plot}. For higher $\epsilon_{A}$ value the particle is distributed with an exponential tail at short time and position limits due to the fact that active diffusivity favors the exponential distribution over normal as depicted in Fig. \ref{p(x)Poisson plot}(a).
		In case of small $\epsilon_{A}$ value the distribution is mostly Gaussian
		at short time and displacement region as pictorially illustrated in Fig.
		\ref{p(x)Poisson plot}(b).  The similarity of these to the experimentally observed probability distributions in \cite{PhysRevLett.103.198103} is very noticeable.

\begin{figure}
\includegraphics[width=0.8\linewidth]{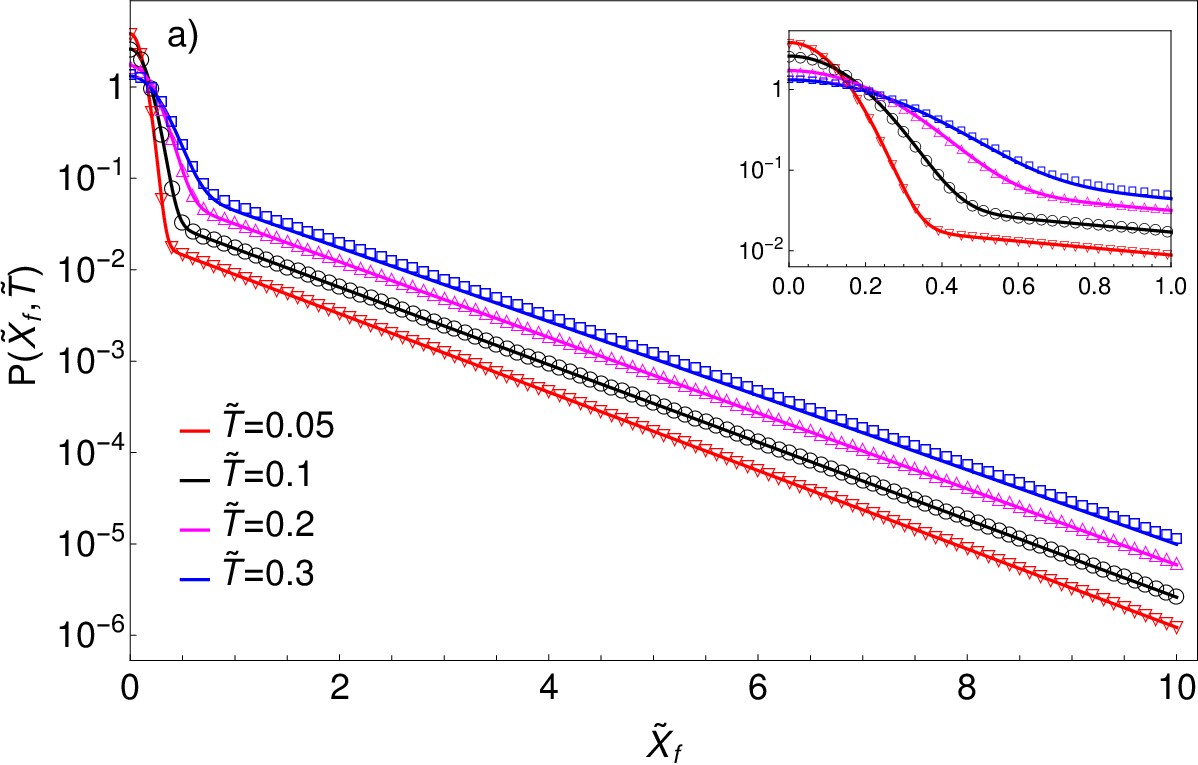} \quad{}\includegraphics[width=0.8\linewidth]{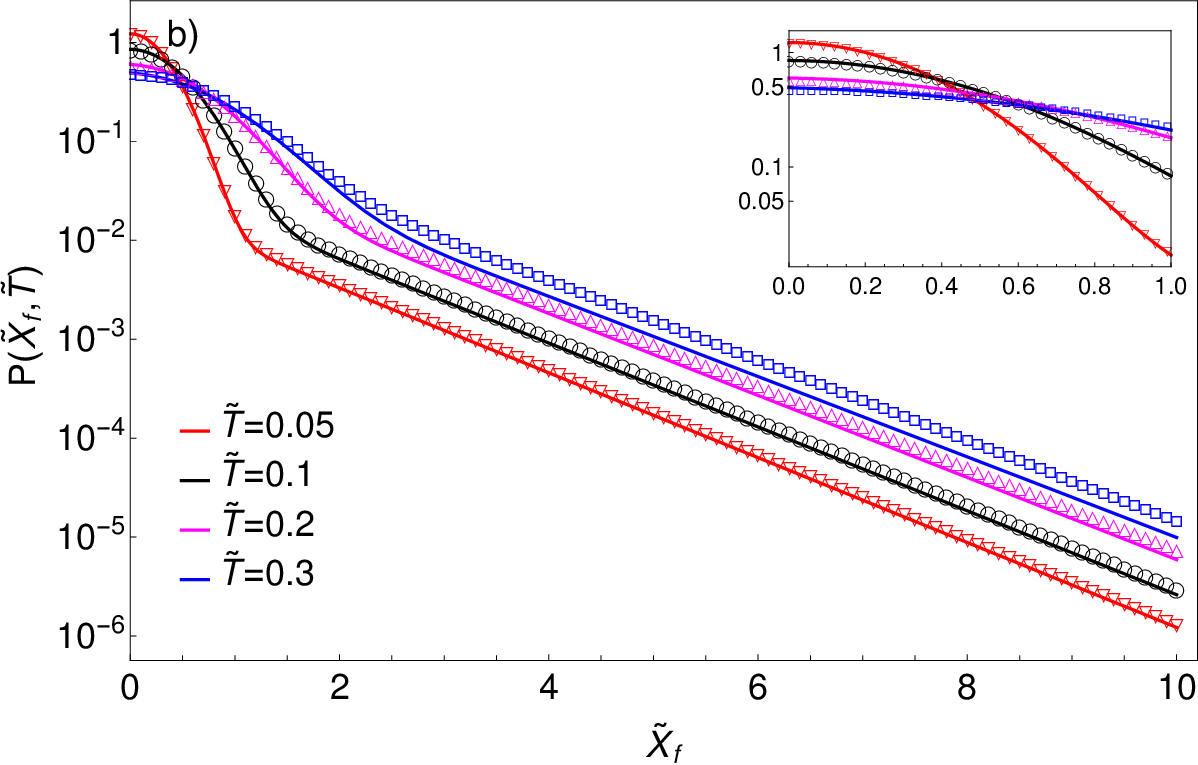}
\caption{Gaussian+Poissonian white noises:  logarithmic values of the probability distribution function are plotted against dimensionless displacement $\tilde{x}_f$ at different scaled times $\tilde{T}$ for two diffusivity ratios (a) $\epsilon_{A}=10$ and (b) $\epsilon_{A}=1.0$. The symbols are used for numerical values obtained after performing integration in Eq. (\ref{rewriting}) numerically. The solid curves represent Eq. (\ref{p(x)pwnsmallfin}) which is an analytical approximation valid at small time scale.  From the plots it is seen that the approximation is quite good for $\tilde{T}\leq0.1$.  The existence of a Gaussian distribution at short $\tilde{x}_f$ and an exponential one for large $\tilde{x}_f$  is clearly seen. }
\label{p(x)Poisson plot} 
\end{figure}
	
\subsubsection{Long time limit}

To get PDF at long time limit, using the identity 
\[
\left(\frac{\tilde{T}}{\tau_{A}(1+\widetilde{p}_{T}^{2})}\right)^{n}=\left(\frac{\tilde{T}}{\tau_{A}}\right)^{n}\frac{1}{(n-1)!}\left\{ \left(-\frac{\partial}{\partial b}\right)^{n-1}\int_{0}^{\infty}d\alpha e^{-\alpha(b+\tilde{p}_{T}^{2})}\right\} _{b=1}
\]
we can write 
\begin{align*}
Q & =\frac{1}{2\pi}\frac{\tilde{T}}{\tau_{A}}\int_{-\infty}^{+\infty}d\widetilde{p}_{T}\,e^{-i\widetilde{p}_{T}\tilde{x}{}_{f}}\,e^{-\frac{\tilde{T}\,\widetilde{p}_{T}^{2}}{\epsilon_{A}}-\tilde{T}}\\
 & \times\sum_{n=0}^{\infty}\frac{\tilde{T}^{n}}{(n+1)!n!}\left\{ \left(-\frac{\partial}{\partial b}\right)^{n}\int_{0}^{\infty}d\alpha e^{-\alpha(b+\widetilde{p}_{T}^{2})}\right\} _{b=1}.
\end{align*}
This may be rewritten as 
\begin{align*}
Q & =e^{-\tilde{T}}\tilde{T}\int d\alpha\sum_{n=0}^{\infty}\frac{\alpha^{n}}{(n+1)!n!}\left(\tilde{T}\right)^{n}e^{-\alpha}\\
 & \text{}\times\frac{1}{2\pi}\int_{-\infty}^{\infty}d\widetilde{p}_{T}\;\;e^{-i\widetilde{p}_{T}\tilde{x}_{f}}\,e^{-(\frac{\tilde{T}}{\epsilon_{A}}+\alpha)\,\widetilde{p}_{T}^{2}}.
\end{align*}
On performing the sum and the integral over $\tilde{p}_{T}$, we get
\[
Q=e^{-\tilde{T}}\tilde{T}\int_{0}^{\infty}d\alpha\;e^{-\alpha}\sqrt{\frac{1}{\tilde{T}\alpha}}I_{1}\left(2\sqrt{\alpha\tilde{T}}\right)\frac{e^{-\frac{x_{f}^{2}}{4(\tilde{T}/\epsilon_{A}+\alpha)}}}{\sqrt{4\pi(\tilde{T}/\epsilon_{A}+\alpha)}}.
\]
 
Changing the variable $\alpha$ to $z$ defined by $\alpha=\tilde{T}z^{2}$
we can write the above as 
\[
Q=e^{-\tilde{T}}2\tilde{T}\int_{0}^{\infty}dze^{-\tilde{T}z^{2}}I_{1}(2\tilde{T}z)\frac{e^{-\frac{x_{f}^{2}}{4\tilde{T}(1/\epsilon_{A}+z^{2})}}}{\sqrt{4\pi\tilde{T}(1/\epsilon_{A}+z^{2})}}.
\]
Using this in Eq. (\ref{rewriting}) gives 
\begin{align}
\mathbb{P}(\tilde{x}_{f},\tilde{T};\,\tilde{x}_{0} & =0,0)=\sqrt{\frac{\epsilon_{A}}{4\pi\tilde{T}}}e^{-\tilde{T}-\epsilon_{A}\tilde{x_{f}^{2}}/(4\tilde{T})}\nonumber \\
 & +e^{-\tilde{T}}2\tilde{T}\int dze^{-\tilde{T}z^{2}}I_{1}(2\tilde{T}z)\frac{e^{-\frac{\tilde{x}_{f}^{2}}{4\tilde{T}(1/\epsilon_{A}+z^{2})}}}{\sqrt{4\pi\tilde{T}(1/\epsilon_{A}+z^{2})}}\label{finalprob.dist.}.
\end{align}
Note that the probability distribution is explicitly written as the
sum (integral) of a collection of Gaussians. Each Gaussian has a width
proportional to $\tilde{T}$. Hence at any instant, the width of the
distribution is proportional to $\tilde{T}$ - implying that Fick's
law is obeyed at all times. For large values of $\tilde{T}$ one can
neglect the first term, and can approximate $I_{1}(2\tilde{T}z)$
in the second term as $\approx e^{2\tilde{T}z}\sqrt{1/(4\pi\tilde{T}z)}$
. This gives 
\[
\mathbb{P}(\tilde{x}_{f},\tilde{T};\,x_{0}=0,0)=\sqrt{\frac{\tilde{T}}{\pi}}\int_{0}^{\infty}dze^{-\tilde{T}(z-1)^{2}}\frac{e^{-\frac{x_{f}^{2}}{4\tilde{T}(1/\epsilon_{A}+z^{2})}}}{\sqrt{4\pi\tilde{T}(1/\epsilon_{A}+z^{2})}}
\]
$e^{-\tilde{T}(z-1)^{2}}$ has its maximum value at $z=1.$ In the
limit $\tilde{T}\rightarrow\infty,$ one can hence replace $z$ everywhere
else by unity, and extend the lower limit of integration to $-\infty.$
This gives

\begin{align}
\mathbb{P}(\tilde{x}_{f},\tilde{T};\,x_{0}=0,0)&=\frac{e^{-\frac{\tilde{x}_{f}^{2}}{4\tilde{T}(1+1/\epsilon_{A})}}}{\sqrt{4\pi\tilde{T}(1+1/\epsilon_{A})}}\sqrt{\frac{\tilde{T}}{\pi}}\int_{-\infty}^{\infty}d\xi\,e^{-\tilde{T}\xi^{2}}\nonumber\\
&=\frac{e^{-\frac{\tilde{x}_{f}^{2}}{4\tilde{T}(1+1/\epsilon_{A})}}}{\sqrt{4\pi\tilde{T}(1+1/\epsilon_{A})}}.
\end{align}
Hence in the long time limit, the result is a Gaussian with $<\tilde{x}_{f}^{2}>=2(1+1/\epsilon_{A})\tilde{T}$,
as one would expect.

\subsection{ $\sigma(t)$ is taken as the position coordinate of an overdamped
harmonic oscillator driven by Poissonian white noise. }
\label{PSNsection}
In this section we will take the active noise $\sigma(t)$ to be correlated
Poissonian, which we will denote by the symbol $\sigma_{PC}(t)$.
As the generation of active force usually happens
as a result of a Poisson process, we assume the active force $\sigma_{PC}(t)$
as the position coordinate of a harmonically bound fictitious particle
driven by Poisson white noise $\sigma_{PW}(t)$. Hence its dynamics
follows 
\begin{equation}
\frac{d\sigma_{PC}(t)}{dt}=-\frac{1}{\tau_p}\sigma_{PC}(t)+\frac{1}{\tau_p}\sigma_{PW}(t)\label{eq:active force dynamics},
\end{equation}
where $\sigma_{PW}(t)$ has been defined in the previous section. We have shown in Appendix \ref{appen3}  that $\sigma_{PC}(t)$ has exponential correlation (see  Eq. (\ref{correlationsigmapcfinal})) and the associated timescale is correlation time, denoted as $\tau_p$.
We can write the characteristic functional of $\sigma_{PC}(t)$ in
terms of Poisson noise $\sigma_{PW}(t)$ as (for details see Appendix \ref{appen3})
\begin{equation}
\left<e^{ip_{T}\int_{0}^{T}\,\sigma_{PC}(t)dt}\right>=\left<e^{i\int_{-\infty}^{+\infty}q(s)\,\sigma_{PW}(s)\,ds}\right>_{\sigma_{PW}(s)},\label{eq:ch zeta(t)}
\end{equation}
where 
\begin{equation}
q(s)=p_{T}[\Theta(-s)\,e^{\lambda_p s}(1-e^{-\lambda_p T})+\Theta(s)\Theta(T-s)\,(1-e^{-\lambda_p(T-s)})]\label{eq:q(s)}
\end{equation}
with $\lambda_p=1/\tau_p$.    $\Theta(t)$ is usual Heaviside theta function.
Using Eq. (\ref{eq:ch functional Poisson}) in Eq. (\ref{eq:ch zeta(t)})
one can obtain the characteristic function  
\begin{align}
\left<e^{ip_{T}\int_{0}^{T}\sigma_{PC}(t)dt}\right>_{\sigma_{PW}(t)}=&\text{exp}\left[-\frac{\tau_{p}}{\tau_{a}}\frac{a_{0}p_{T}}{1+a_{0}^{2}p_{T}^{2}}\left(a_{0} \lambda_p T p_{T}-\text{tan}^{-1}(a_{0}\rho p_{T})\right)\right]\nonumber\\
&\times\,(1+a_{0}^{2}\rho^{2}p_{T}^{2})^{-\frac{\tau_{p}}{2\tau_{a}}\frac{a_{0}^{2}p_{T}^{2}}{1+a_{0}^{2}p_{T}^{2}}}.\label{eq:final ch active}
\end{align}
Here, $\rho=1-e^{-T\lambda_p}$ and $\tau_{a}$ is inverse of Poisson rate $\mu$ $(\text{i.e.,}\:\tau_{a}=\frac{1}{\mu})$, and is the 
characteristic timescale for the active noise. $a_{0}$ is the characteristic
length scale of the active noise. We define the diffusivity due to the active
noise as $D_{A}=\frac{a_{0}^{2}}{\tau_{a}}$.

Using Eq. (\ref{eq:final ch active}) in Eq. (\ref{eq:finP3}),
leads to the probability distribution function for finding the particle
at the position $x_{f}$ after a time $T$,  given that it started at $x_{0}=0$ at time
$T=0$ as 
\begin{equation}
\mathbb{P}(x_{f},T;\,x_{0}=0,0)=\frac{1}{2\pi}\int_{-\infty}^{+\infty}dp_{T}\,e^{-ip_{T}x_{f}}\mathbb{F}(p_{T})\label{eq:final P active},
\end{equation}
where 
\begin{equation}
\mathbb{F}(p_{T})=e^{-D\,T\,p_{T}^{2}}\,\text{exp}\left[-\frac{\tau_{p}}{\tau_{a}}\frac{a_{0}p_{T}}{1+a_{0}^{2}p_{T}^{2}}
\left(a_{0} \lambda_p T  p_{T}-\text{tan}^{-1}(a_{0}\rho p_{T})\right)\right](1+a_{0}^{2}\rho^{2}p_{T}^{2})^{-\frac{\tau_{p}}{2\tau_{a}}\frac{a_{0}^{2}p_{T}^{2}}{1+a_{0}^{2}p_{T}^{2}}}.\label{fin F active }
\end{equation}
Using Eq. (\ref{m-th moment}), the MSD is calculated from Eq. (\ref{fin F active })
to be 
\begin{equation}
<x(T)^{2}>=2DT+2\frac{D_{A}}{\lambda_p}\left[T \lambda_p+e^{-T\lambda_p}-1\right].\label{msd active}
\end{equation}
In the small time limit ($\lambda_p T\ll 1$), $\left<x(T)^{2}\right>=2DT$ suggesting that
the motion of the particle is influenced by only white noise. At the intermediate time scale $i.e.$ when $0<\lambda_p T <1$, it starts
to feel the effect of active noise.  In the large time limit ($\lambda_p T \gg 1$) it diffuses
with an enhanced diffusivity $D+D_{A}$ as the MSD is given by
$\left<x(T)^{2}\right>=2(D+D_{A})T$.

To get an inkling about the behavior of PDF at large displacement
we have computed $\gamma_{np}$ applying Eq. (\ref{NGP}) and it is
given by
\begin{align}
\gamma_{np}(T)&=\frac{D_{A}\,a_{0}^{2}[-11+6T\lambda_p +18e^{-T\lambda_p }-9e^{-2T\lambda_p }+2e^{-3T\lambda_p }]}{3\tau_{p}[D_{A}(-1+T\lambda_p +e^{-T\lambda_p })+D\,T\lambda_p ]^{2}}\nonumber\\
&=\frac{\tilde{\tau}\,\epsilon_A^2}{3}\frac{[-11+6 \tilde{T}+18e^{-\tilde{T}}-9e^{-2 \tilde{T}}+2e^{-3 \tilde{T}}]}{[\epsilon_{A}(-1+\tilde{T}+e^{-\tilde{T}})+\tilde{T}]^{2}}\label{NGP active}.
\end{align}
Here we have expressed the $\gamma_{np}(\tilde{T})$ in terms of the dimensionless variables defined  by $\tilde{T}=\frac{T}{\tau_p}=\lambda_p T,\,\tilde{\tau}=\frac{\tau_a}{\tau_p},\,\epsilon_A=\frac{D_A}{D}.$  At small and large times, $\gamma_{np}(T)$ vanishes indicating 
Gaussian behavior of the probability distribution. At intermediate time scale, $\gamma_{np}(T)$ attains a maximum  and then it slowly decays to zero in the
long time limit as shown in Fig. \ref{ngpactive plot}.  This  suggests
that the distribution deviates from Gaussianity significantly in the intermediate time range as  before the particle samples around the local surroundings it is triggered by the active noise. As $\epsilon_A$ takes higher value,  the peak height keeps increasing indicating  that the active noise is responsible for the non-Gaussianity.
\begin{figure}[h]
\includegraphics[width=0.8\linewidth]{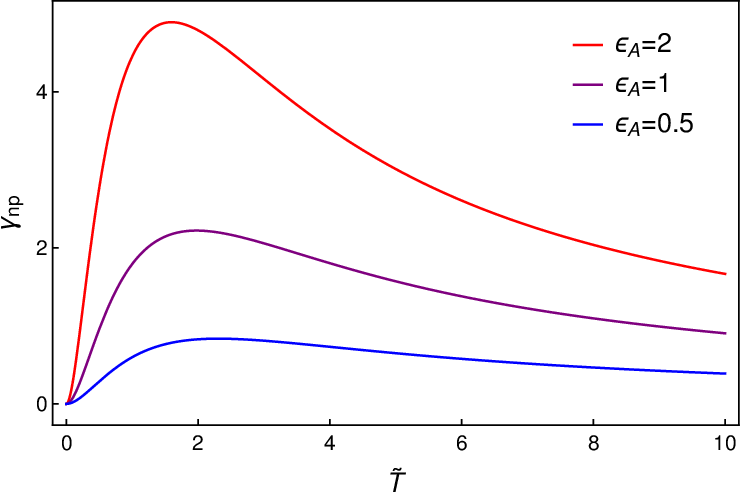}
\caption{ $\sigma_{PC}(t)$+Gaussian white noise: non-Gaussian parameter $(\gamma_{np})$ is plotted as a function of scaled time $\tilde{T}$ for three different values of $\epsilon_{A}$, keeping $\tilde{\tau} (=\tau_a/\tau_p)$ fixed at $20$. The features of the plots remain unaltered for any particular ratio of $\tau_{p}$ and $\tau_{a}$.}\label{ngpactive plot}
\end{figure}
Now the complete features of the dynamics can be investigated by finding PDF. To do that we have to invert $\mathbb{F}(p_{T})$ to the position space as shown in Eq. (\ref{eq:final P active}). But as the inverse Fourier transform cannot be done analytically, one needs to do it numerically.   However, one can get the asymptotic behaviors of the PDF. Before proceeding further we rescale $x_f$, $p_T$ by $a_0$, introducing $\tilde{x}_f=\frac{x_f}{a_0}$, $\tilde{p_T}=a_0\,p_T.$  Eq. (\ref{eq:final P active}) can be rewritten in terms of these as 
\begin{equation}
\mathbb{P}(\tilde{x}_{f},\tilde{T};\,\tilde{x}_{0}=0,0)=\frac{1}{2\pi}\int_{-\infty}^{+\infty}d\tilde{p}_{T}\,e^{-i\tilde{p}_{T}\tilde{x}_{f}}\mathbb{F}(\tilde{p}_{T})\label{final P' active},
\end{equation}
where 
\begin{equation}
\mathbb{F}(\tilde{p}_{T})=e^{-\frac{\tilde{T} }{\tilde{\tau}}\frac{\tilde{p}_T^2}{\epsilon_A}}\,\text{exp}\left[-\frac{1}{\tilde{\tau}}\frac{\tilde{p}_{T}}{1+\tilde{p}_{T}^{2}}\left(\tilde{T}\,\tilde{p}_{T}-\text{tan}^{-1}(\rho\,\tilde{p}_{T})\right)\right](1+\rho^{2}\tilde{p}_{T}^{2})^{-\frac{1}{2 \tilde{\tau}}\frac{\tilde{p}_{T}^{2}}{1+\tilde{p}_T^{2}}}\label{finF'active}.
\end{equation}
We now consider different limits, which can be looked at analytically.
\subsubsection{$\tilde{T}\ll 1$ and $\tilde{x}_{f}$ arbitrary.}
In the short time limit,  
\begin{equation} \exp  \left[-\frac{1}{\tilde{\tau}}\frac{\tilde{p}_{T}}{1+\tilde{p}_{T}^{2}}\left(\tilde{T}\,\tilde{p}_{T}-\tan^{-1}(\rho\,\tilde{p}_{T})\right)\right]\approx 1,  \end{equation} so that 
\begin{equation}
\mathbb{F}(\tilde{p}_{T})\approx e^{-\frac{\tilde{T} }{\tilde{\tau}}\frac{\tilde{p}_T^2}{\epsilon_A}}\,(1+\rho^{2}\tilde{p}_{T}^{2})^{-\frac{1}{2 \tilde{\tau}}\frac{\tilde{p}_{T}^{2}}{1+\tilde{p}_T^{2}}}\label{finF'active}.
\end{equation}
So the probability distribution function is the convolution of a Gaussian, of width $\sim \sqrt{\tilde{T}/(\tilde{\tau}\epsilon_{A})}$ with the inverse Fourier transform of the function $(1+\rho^{2}\tilde{p}_{T}^{2})^{-\frac{1}{2 \tilde{\tau}}\frac{\tilde{p}_{T}^{2}}{1+\tilde{p}_T^{2}}}$, which we will denote as $h(\tilde{x}_{f})$.  Though the inverse Fourier transform cannot be done analytically, the width of the resulting function can be estimated to be $\sim \rho\approx \tilde{T}$.   As $\tilde{T}\ll 1$, if one has $\sqrt{\tilde{T}/(\tilde{\tau}\epsilon_{A})} \gg \tilde{T}$ then $h(\tilde{x}_{f}) $ can be approximated as a Dirac delta function.  Hence in this limit
\begin{equation}
\mathbb{P}(\tilde{x}_{f},\tilde{T};\,\tilde{x}_{0}=0,0)=\sqrt{\frac{\epsilon_{A}\tilde{\tau}}{4\pi \tilde{T}}}\exp \left ( -\frac{\epsilon_{A}\tilde{\tau}\tilde{x}_{f}^{2}}{4\tilde{T}} \right ). \label{KLSresult} 
\end{equation}
Thus the distribution displays predominant signature of white noise.
\begin{figure}[h!]
	\includegraphics[width=0.8\linewidth]{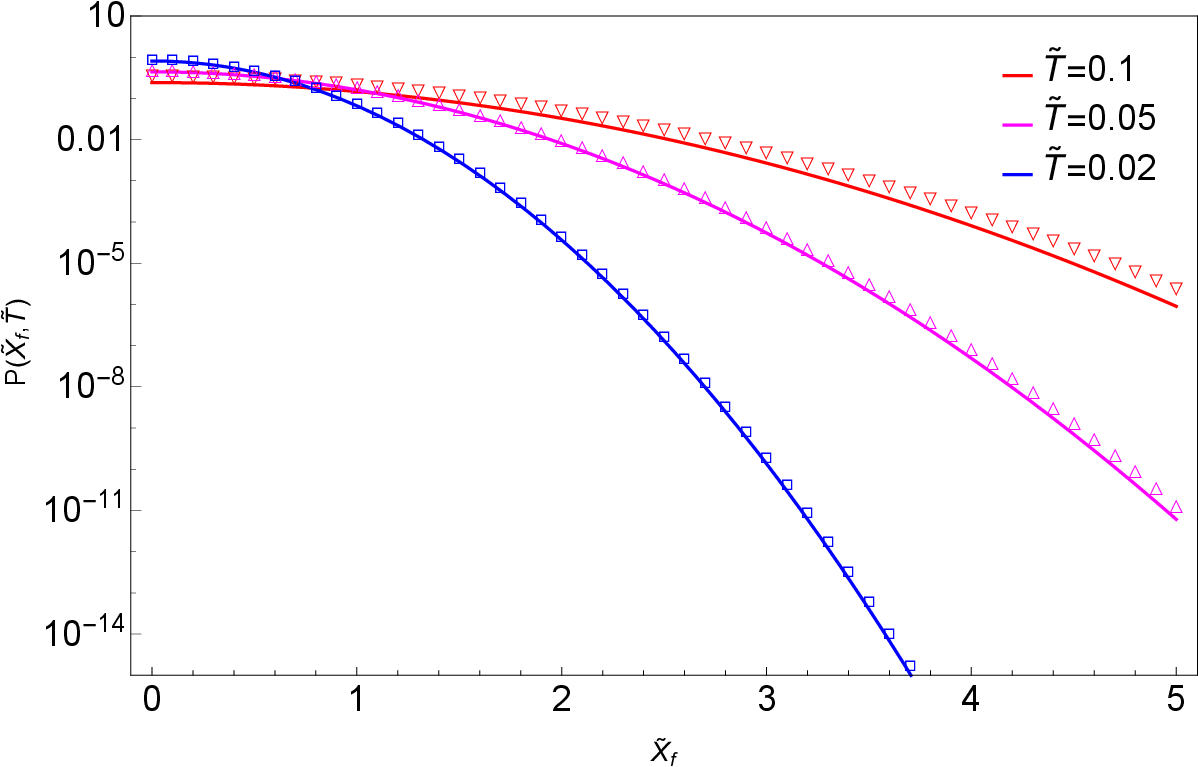} \caption{  Probability distribution function,   plotted as a function of  the scaled displacement in logarithmic scale  for different values of time.  The plots are for the values of parameters: $\{\tilde{\tau}=0.2, \epsilon_A=1\}$. The plots with symbols are obtained after numerical integration  of  Eq. (\ref{final P' active}) and solid curves represent Eq. (\ref{KLSresult}). }
	\label{P(x)plot_smalltau} 
\end{figure}
\subsubsection{Long time ($\tilde{T}\gg 1$), large $\tilde{x}_{f}$ ($\tilde{x}_{f}\gg 1 $) limit. }
At large displacement ($i.e.$ $x_f\gg a_0$, or equivalently $\tilde{p}_{T}\ll 1$) and for long time ($ \tilde{T}\gg 1,\; \Rightarrow \rho  \approx 1$) limit one can express $\mathbb{F}(\tilde{p}_{T})$ as
a Gaussian function of $\tilde{p}_{T}$, $viz.$,
\begin{equation}
\mathbb{F}(\tilde{p}_{T})\approx e^{-\frac{\tilde{T}}{\tilde{\tau}}\left(1+\frac{1}{\epsilon_A}\right)\tilde{p}_T^2}.\label{active ch large x}
\end{equation}
Hence the PDF can be easily obtained by Fourier inversion of Eq. (\ref{active ch large x}) and it is given as
\begin{equation}
\mathbb{P}(\tilde{x}_{f},\tilde{T};\,\tilde{x}_{0}=0,0)=\sqrt{\frac{\tilde{\tau}\,\epsilon_A}{4(1+\epsilon_A)\,\pi \tilde{T}}}\,e^{-\frac{\epsilon_A}{1+\epsilon_A}\frac{\tilde{\tau}\,\tilde{x}_f^2}{4 \tilde{T}}}\label{active ch large x,longt pdf}.
\end{equation}
So, in this limit the particle is Gaussian-distributed with a total diffusivity of $D+D_{A}$.

\subsubsection{Different time limits, but small $\tilde{x}{}_{f}$  }
 In this case,   $\tilde{x}_f \ll 1.$
  This implies
  $\tilde{p}_{T}\gg 1$, and the terms in Eq. (\ref{finF'active}) can be approximated as follows:
 $$\text{exp}\left[-\frac{1}{\tilde{\tau}}\frac{\tilde{p}_{T}}{1+\tilde{p}_{T}^{2}}\left(\tilde{T}\,\tilde{p}_{T}-\text{tan}^{-1}(\rho\, \tilde{p}_{T})\right)\right]\approx \text{exp}\left[-\frac{\tilde{T}}{\tilde{\tau}}+\frac{\tilde{T}}{\tilde{p}_T}\text{tan}^{-1}(\rho\, \tilde{p}_{T})\right] \approx e^{-\frac{\tilde{T}}{\tilde{\tau}}},$$
 and
 \begin{align}
 (1+\rho^{2}\tilde{p}_{T}^{2})^{-\frac{1}{2\tilde{\tau}}\frac{\tilde{p}_{T}^{2}}{1+\tilde{p}_T^{2}}}\approx (1+\rho^{2}\tilde{p}_{T}^{2})^{-\frac{1}{2\tilde{\tau}}}\label{smallx_int}.
 \end{align}

Therefore, distribution of displacement as given in Eq. (\ref{final P' active}) can be approximated as 
\begin{align}
\mathbb{P}(\tilde{x}_{f},\tilde{T};\,\tilde{x}_{0}=0,0)\approx\frac{e^{-\frac{\tilde{T}}{\tilde{\tau}}}}{2\pi}\int_{-\infty}^{+\infty}d\tilde{p}_{T}\,e^{-i\tilde{p}_{T}\tilde{x}_{f}}\frac{ e^{-\frac{\tilde{T}}{\tilde{\tau}}\frac{ \tilde{p}_T^2}{\epsilon_A}}}{(1+\rho^{2}\tilde{p}_{T}^{2})^{\frac{1}{2\tilde{\tau}}}}\label{pcn_p_int}.
\end{align}
The above can be rewritten as a convolution of two functions as outlined in section \ref{PWNsection} and it reads
\begin{align}
\mathbb{P}(\tilde{x}_{f},\tilde{T};\,\tilde{x}_{0}=0,0)=\int_{-\infty}^{+\infty} d\tilde{x}\,\sqrt{\frac{\epsilon_{A}\,\tilde{\tau}}{4\pi\tilde{T}}}e^{-\frac{\tilde{T}}{\tilde{\tau}}-\epsilon_{A}\,\tilde{\tau}(\tilde{x}_{f}-\tilde{x})^{2}/(4\tilde{T})}\,\frac{2^{\frac{3}{2}-\frac{1}{2\tilde{\tau}}}\,\sqrt{\pi}\,\left|\frac{\widetilde{x}}{\rho}\right|^{\frac{1}{2\tilde{\tau}}-\frac{1}{2}}K_{\frac{1}{2\tilde{\tau}}-\frac{1}{2}}\left(\frac{\left|\widetilde{x}\right|}{\rho}\right)}{\Gamma(\frac{1}{2\tilde{\tau}})}\label{P_pcn_int1}.
\end{align} 
The above integration cannot be done  analytically. So we analyze limiting situations in the following. In the limit where such analysis is not possible, we evaluate the integral numerically.  Our numerical results match these analytical results, and also interpolate between these limits. 

 The Gaussian, due to thermal noise has the  width $\sqrt{\frac{2\tilde{T}}{\epsilon_A\tilde{\tau}}}$ and the Bessel function has the width $\rho$. If the Gaussian has negligible width compared to that of  Bessel function, $i.e.$, if $\frac{\epsilon_A \tilde{\tau}\rho^2}{2\tilde{T}}\gg \rho \approx 1$, then the Gaussian can be approximated as a delta function, and the integration performed. The result is
\begin{equation}
\mathbb{P}_{}(\tilde{x}_{f},\tilde{T})\approx
\frac{2^{\frac{3}{2}-\frac{1}{2\tilde{\tau}}}\,\sqrt{\pi}}{\Gamma(\frac{1}{2\tilde{\tau}})}
 \left|\frac{\widetilde{x}_{f}}{\rho}\right|^{\frac{1}{2\tilde{\tau}}-\frac{1}{2}}K_{\frac{1}{2\tilde{\tau}}-\frac{1}{2}}\left(\frac{\left|\widetilde{x}_{f}\right|}{\rho}\right) \label{bessel_pcn}.
\end{equation}
On the other hand, in the limit $\sqrt{\frac{2\tilde{T}}{\epsilon_A\tilde{\tau}}} \gg \rho$  or for $1\gg \frac{\epsilon_A\tilde{\tau}\rho^2}{2\tilde{T}}$,  Bessel function can be approximated by a Dirac delta function and one gets
\begin{align}
\mathbb{P}(\tilde{x}_{f},\tilde{T};\,\tilde{x}_{0}=0,0) \approx e^{-\frac{\tilde{T}}{\tilde{\tau}}}\sqrt{\frac{\epsilon_A\tilde{\tau}}{4\pi\tilde{T}}}e^{-\frac{\epsilon_A\tilde{\tau}}{4\tilde{T}}\tilde{x}_f^2}\label{smallx_smalltau}.
\end{align}

 Another approximation, which requires $\rho \tilde{p}_T \gg 1$ and $\tilde{p}_T \gg 1$ is given below.  In this case, 
\[\mathbb{F}(\tilde{p}_{T}) \approx    \frac{e^{-\frac{\tilde{T}}{\tilde{\tau}}}}{2\pi}\int_{-\infty}^{+\infty}d\tilde{p}_{T}\,e^{-i\tilde{p}_{T}\tilde{x}_{f}}\frac{ e^{-\frac{\tilde{T}}{\tilde{\tau}}\frac{ \tilde{p}_T^2}{\epsilon_A}}}{(1+\rho^{2}\tilde{p}_{T}^{2})^{\frac{1}{2\tilde{\tau}}}},      \]
which may be approximated as
	\begin{align}
	\mathbb{P}(\tilde{x}_{f},\tilde{T};\,\tilde{x}_{0}=0,0)\approx\frac{e^{-\frac{\tilde{T}}{\tilde{\tau}}}}{2\pi}\int_{-\infty}^{+\infty}d\tilde{p}_{T}\,e^{-i\tilde{p}_{T}\tilde{x}_{f}}\frac{ e^{-\frac{\tilde{T}}{\tilde{\tau}}\frac{ \tilde{p}_T^2}{\epsilon_A}}}{(\rho^{2}\tilde{p}_{T}^{2})^{\frac{1}{2\tilde{\tau}}}}\label{pcn_p_int1}.
	\end{align}  
For $\tilde{\tau}>1,$ the above can be done analytically and leads to
\begin{align}
\mathbb{P}(\tilde{x}_f)=\frac{\Gamma\left(\frac{1}{2}-\frac{1}{2\,\tilde{\tau}}\right)}{2 \pi}\,e^{-\frac{\tilde{T}}{\tilde{\tau}}}\rho^{-\frac{1}{\tilde{\tau}}}\,\left(\frac{\tilde{T}}{\tilde{\tau}\,\epsilon_a}\right)^{-\frac{1}{2}(1-\frac{1}{\tilde{\tau}})}{}_{1}F_{1}\left(\frac{1}{2}-\frac{1}{2\,\tilde{\tau}};\frac{1}{2};-\epsilon_A\,\frac{\tilde{x}_{f}^{2}\,\tilde{\tau}}{4\tilde{T}}\right)\label{active approx int x},
\end{align} 
where $_{1}F_{1}(a,b;z)$ is Kummer confluent hypergeometric function of first
kind \cite{olver2010nist}. The function in Eq. (\ref{active approx int x}) has a central Gaussian core like  Eq. (\ref{smallx_smalltau}), and deviates from it as may be seen in the insets of  Fig. \ref{pdf_pcnplot_times}.

\begin{figure}[H]
	\includegraphics[width=0.8\linewidth]{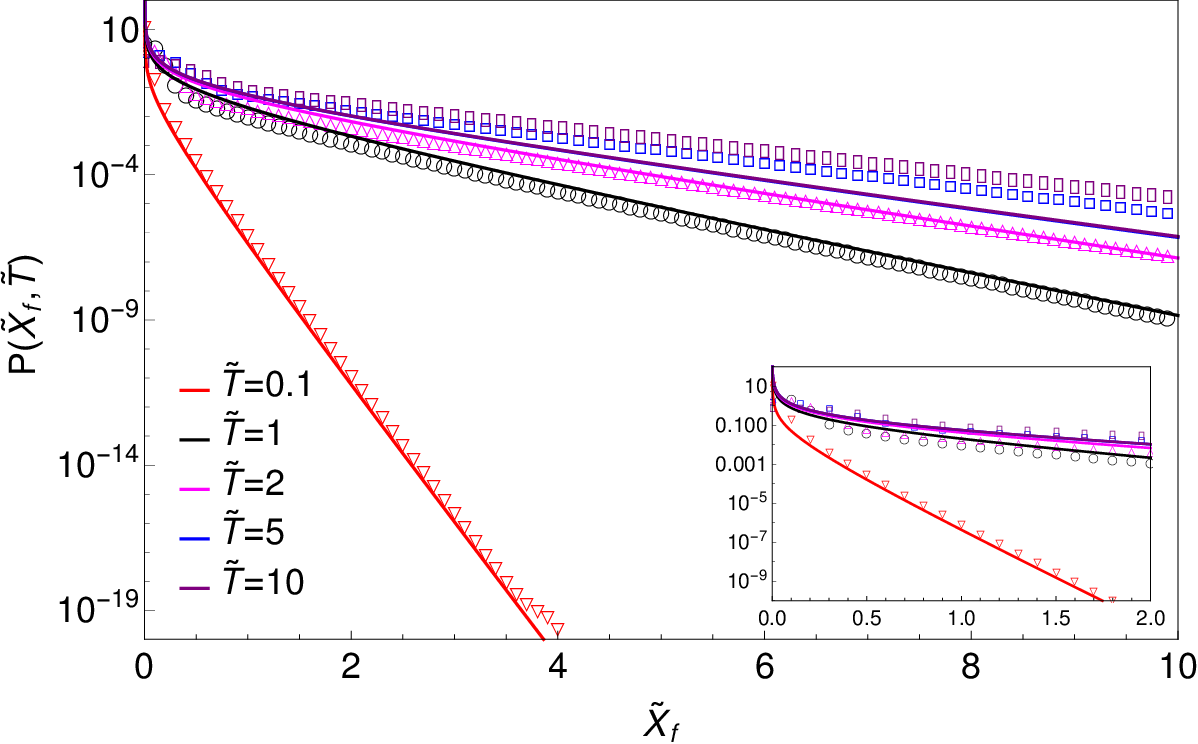} \caption{Logarithmic values of probability distribution function  is plotted against scaled displacement for different values of time keeping  parameters fixed at $\tilde{\tau}=20$, $\epsilon_A=10$. The plots with symbols is obtained from Eq. (\ref{final P' active}) after numerical integration and solid curves represent Eq. (\ref{bessel_pcn}). The same plots have been enlarged in the inset to obtain behavior  near  origin.  }
	\label{P(x)plot} 
\end{figure}

 We have performed the numerical integration of Eq. (\ref{final P' active}) and made plots at different times for the same normal and active diffusivities,  as shown in Fig. \ref{P(x)plot_smalltau}- \ref{pdf_pcnplot_times}. Fig. \ref{P(x)plot_smalltau} demonstrates the  behavior of distribution in the limit $1\gg \frac{\epsilon_A\tilde{\tau}\rho^2}{2\tilde{T}}$ at small position  and suggests Gaussian distribution of the form given in  Eq. (\ref{KLSresult}). Fig. \ref{pdf_pcnplot_times} is drawn in the following regime : $\frac{\tilde{T}}{\tilde{\tau}}\ll 1$, $\tilde{\tau}>1$ and $\epsilon_A \sim\mathcal{O}(1)$. In the inset,  numerical values show that at small $\tilde{x}_{f}$, the distribution is Gaussian, but at higher values it deviates.  The deviation from Gaussianity is described by Eq. (\ref{bessel_pcn}).     For the system with higher active diffusivity, the central Gaussian shape has virtually zero width. Hence, throughout entire position range it basically follows the same distribution (\ref{bessel_pcn}) as shown in Fig. \ref{P(x)plot}.  
 
\begin{figure}[H]
\includegraphics[width=0.8\linewidth]{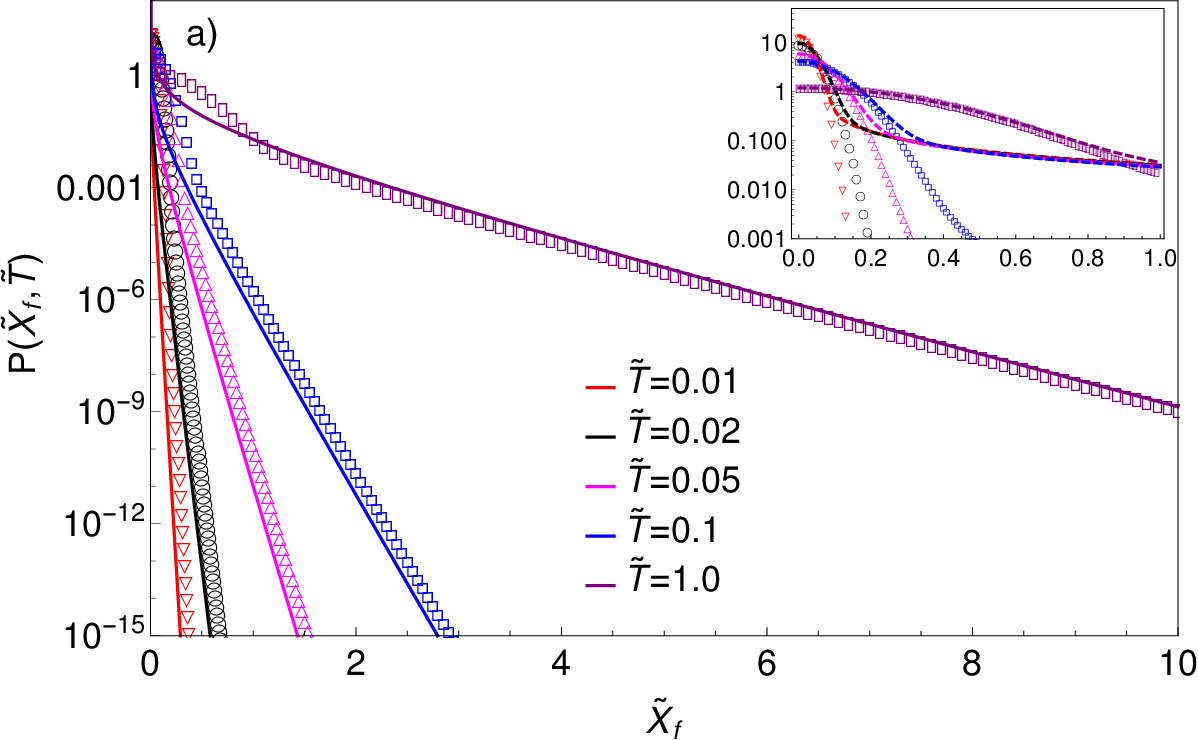} \quad{}\\
\includegraphics[width=0.8\linewidth]{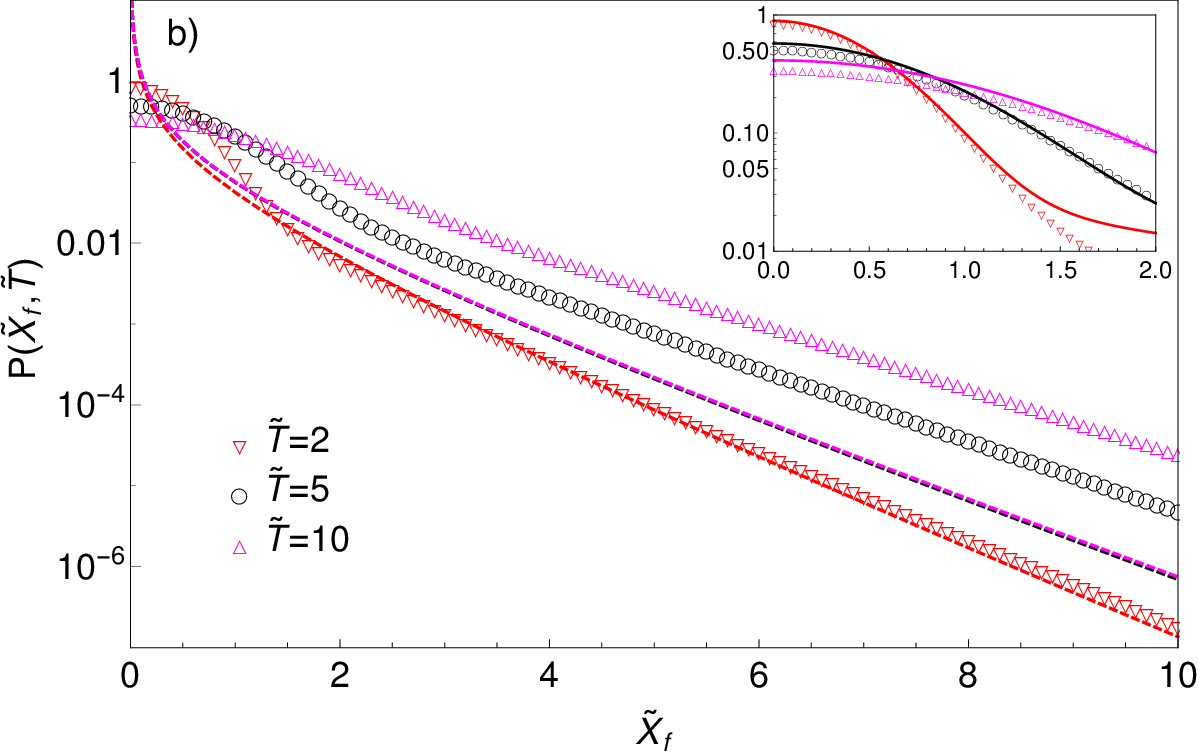}
\caption{The Logarithmic plot of probability distribution function vs. displacement  at intermediate time scale for (a) $\tilde{T}\ll \tilde{\tau}$  and (b) $\tilde{T}<\tilde{\tau}$, where $\tilde{\tau}>1$. All the curves with symbols have been obtained from numerical integration of Eq. (\ref{final P' active}) taking $\epsilon_A=1$ and $\tilde{\tau}=20$. In Fig. a, Eq. (\ref{bessel_pcn}) are drawn as solid curves for entire position range at different times, which agree well with the numerical results at positions away from origin. Near the origin, the result agrees well with Eq. (\ref{active approx int x}) corresponding to dashed curves as shown in the inset. In the inset of  b), solid curves correspond to Eq. (\ref{active approx int x}) for different times which are quite a good fit. The results in b) are for longer times.   The curves are obtained using Eq. (\ref{bessel_pcn}) and are seen to be in poor agreement with numerical results.}
\label{pdf_pcnplot_times}
\end{figure}

\section{Results for harmonic oscillator in different types of noises $\sigma(t)$}

\subsection{ Gaussian colored Noise }

Consider $\sigma(t)$ as colored Gaussian noise and the particle is
initially at an equilibrium distribution of the form:  $P(x_{0})=\sqrt{\frac{\lambda}{2\pi D}}\,e^{-\frac{\lambda}{2D}\,x_{0}^{2}}$.
The characteristic function $\mathbb{G}(p_{T})$ in Eq. (\ref{Gsho})
can be computed using Eq. (\ref{ch gaussian}) and it is given as
\begin{equation}
\mathbb{G}(p_{T})=\text{exp}\left[-\frac{D_{A\,}p_{T}^{2}}{2\lambda(\lambda^{2}\tau^{2}-1)}\left((\lambda\tau-1)-2\lambda\tau e^{-T(\lambda+1/\tau)}+(\lambda\tau+1)e^{-2T\lambda}\right)\right].
\end{equation}
Therefore the Fourier transform of probability distribution function,
$\mathbb{F}(p_{T})$ can be written as 
\begin{equation}
\mathbb{F}(p_{T})=\text{exp}\left[-\frac{D_{A\,}p_{T}^{2}}{2\lambda(\lambda^{2}\tau^{2}-1)}\left((\lambda\tau-1)-2\lambda\tau e^{-T(\lambda+1/\tau)}+(\lambda\tau+1)e^{-2T\lambda}\right)-\frac{D\,p_{T}^{2}}{2\lambda}\right].
\end{equation}
Using Eq. (\ref{m-th moment}) one can obtain MSD which is given as
\begin{equation}
<x(T)^{2}>=\frac{D_{A}}{\lambda(\lambda^{2}\tau^{2}-1)}\left((\lambda\tau-1)-2\lambda\tau e^{-T(\lambda+1/\tau)}+(\lambda\tau+1)e^{-2T\lambda}\right)+\frac{D}{\lambda}.
\end{equation}
At short time limit $<x(T)^{2}>=D/\lambda$, which is the MSD at equilibrium
distribution. But with the passage of time, $\sigma(t)$ comes into
effect as it is evident from the value of MSD, which is given for very
large time limit as: $<x(T)^{2}>=\frac{1}{\lambda}\left[D+\frac{D_{A}}{1+\lambda\tau}\right].$

As  $\mathbb{F}(p_{T})$ is Gaussian, its Fourier transform is easily computed to get
\begin{equation}
\mathbb{P}(x_{f},T)=\sqrt{\frac{1}{2\pi\theta^{2}}}\;e^{-\frac{x_{f}^{2}}{2\theta^{2}}},
\end{equation}
where $\theta^{2}=<x(T)^{2}>.$ Therefore the particle is  Gaussian
distributed at all times, with a  width which has contribution from both $D$
and $D_{A}$. At large time limit the width of the distribution becomes
time independent as obvious with a value $\frac{1}{\lambda}\left[D+\frac{D_{A}}{1+\lambda\tau}\right]$.

\subsection{ Dichotomous Poisson Noise}
We consider here the case where the dynamics of a particle ($e.g.$, Janus particles or flagellated bacteria) is driven by both white Gaussian and dichotomous noise, in a harmonic potential $V(x)=\frac{\lambda\,x^2}{2}$. The dichotomous noise  has mean zero and an exponential correlation, $viz.$,
\begin{align}
&\left<\sigma_{DP}(t)\right>=0\nonumber\\
\left<\sigma_{DP}(t)\sigma_{DP}(t')\right>=&u^2 e^{-\gamma \vert t-t' \vert}=D_A\,\gamma\,e^{-\gamma \vert t-t' \vert}\label{correlationdicho}.
\end{align}
Here, diffusivity in active noise $D_A$ is defined as: $D_A=\frac{u^2}{\gamma}$. The noise  $\sigma(t)$ can be described by the characteristic function defined by  \cite{PhysRevA.41.754}
\begin{align}
G(p_T)=&\left<e^{ip_T\,e^{-\lambda  T}\int_{0}^{T}dt\,e^{\lambda t}\sigma_{DP}(t)}\right>={}_{0}F_1\left(\frac{\alpha+1}{2},-\frac{p_T^2 u^2}{4 \lambda^2}\right)\,_{0}F_1\left(\frac{-\alpha+1}{2},-\frac{p_T^2 u^2}{4 \lambda^2}\,e^{-2 \lambda T}\right)\nonumber\\
&+\frac{u^2 p_T^2\,e^{-(\gamma+\lambda)T}}{\lambda^2(\alpha+1)(-\alpha+1)}\,_{0}F_1\left(\frac{-\alpha+3}{2},-\frac{p_T^2 u^2}{4 \lambda^2}\right)\,_{0}F_1\left(\frac{\alpha+3}{2},-\frac{p_T^2 u^2}{4 \lambda^2}\,e^{-2 \lambda T}\right)\label{ch_function_dicho_sho},
\end{align} 
where $\alpha=\frac{\gamma}{\lambda},$ which is the signature of the active noise.
Considering the particle to be initially in the equilibrium state which would result in the absence of active noise $\sigma_{DP}$, the probability distribution function that would result at $T$ may be written as
\begin{align}
\mathbb{P}(x_f,T)= \frac{1}{2 \pi}\int_{-\infty}^{+\infty}dp_T\,e^{-i p_T x_f}\,e^{-\frac{D}{2 \lambda}p_T^2}\,G(p_T)\label{pdf_dichosho}.
\end{align}
 It should be noted that the final steady state form of  the distribution would be independent of the initial starting distribution. Now let us  define dimensionless variables as: $\tilde{x}_f=x_f\frac{\gamma}{u},\,\tilde{p}_T=p_T\frac{u}{\gamma},\,\tilde{T}=\lambda T,\,\epsilon_A=\frac{D_A}{D}.$ Then, Eq. (\ref{pdf_dichosho}) can be rewritten as
\begin{align}
\mathbb{P}(\tilde{x}_f,\tilde{T})= \frac{1}{2 \pi}\int_{-\infty}^{+\infty}d\tilde{p}_T\,e^{-i \tilde{p}_T \tilde{x}_f}\,e^{-\frac{\alpha}{2 \epsilon_A}\tilde{p}_T^2}\,G(\tilde{p}_T)\label{pdf_dichosho_1},
\end{align}
with
\begin{align}
G(\tilde{p}_T)=&_{0}F_1\left(\frac{\alpha+1}{2},-\frac{\tilde{p}_T^2 \alpha^2}{4}\right)\,_{0}F_1\left(\frac{-\alpha+1}{2},-\frac{\tilde{p}_T^2 \alpha^2}{4}\,e^{-2 \tilde{ T}}\right)\nonumber\\
&+\frac{\tilde{p}_T^2\,\alpha^2\,e^{-(\alpha+1)\tilde{T}}}{(\alpha+1)(-\alpha+1)}\,_{0}F_1\left(\frac{-\alpha+3}{2},-\frac{\tilde{p}_T^2 \alpha^2}{4}\right)\,_{0}F_1\left(\frac{\alpha+3}{2},-\frac{\tilde{p}_T^2 \alpha^2}{4}\,e^{-2 \tilde{T}}\right)\label{ch_function_dicho_sho_1}.
\end{align} 
The MSD of the distribution at time $\tilde{T}$ is \begin{align}
\left<\tilde{x}^2(\tilde{T})\right>=\frac{\alpha^2}{1+\alpha}-\frac{2\alpha^2}{1-\alpha^2}e^{-(1+\alpha) \tilde{T}}+\frac{\alpha^2}{1-\alpha}e^{-2 \tilde{T}}+\frac{\alpha}{\epsilon_A}\label{msd_dicho_sho}.
\end{align}
At short time, $\left<\tilde{x}^2(\tilde{T})\right> \sim \frac{\alpha}{\epsilon_A}$ suggesting as appropriate for the initial  equilibrium state. Notice that $\left<\tilde{x}^2(\tilde{T})\right> \rightarrow 0$ if $\epsilon_A \rightarrow \infty$, reflecting the fact that initial contribution comes from white noise. After a long time, $\left<\tilde{x}^2(\tilde{T})\right> \sim \frac{\alpha^2}{1+\alpha}+\frac{\alpha}{\epsilon_A},$ which includes the effect of active noise.  By virtue of Eq. (\ref{m-th moment}) and Eq. (\ref{pdf_dichosho_1}), fourth moment is calculated at $\tilde{T} \rightarrow \infty$ limit and it is given by \begin{align}
	\left<\tilde{x}^4(\tilde{T})\right>=\frac{3 \alpha ^4}{2 (\alpha +1) \left(\frac{\alpha +1}{2}+1\right)}+\frac{6 \alpha ^3}{(\alpha +1) \epsilon _A}+\frac{3 \alpha ^2}{\epsilon _A^2}.
	\end{align} 
	Subsequently, using Eq. (\ref{NGP}) NGP is computed in this limit and it reads
	\begin{align}
	\gamma_{np}=-\left(\frac{2}{3+\alpha}\right)\left(\frac{1}{1+\frac{1}{\epsilon_A}+\frac{1}{\alpha\,\epsilon_A}}\right)^2.
	\end{align}

For the limit $\alpha \rightarrow \infty$, $\gamma_{np}$ diminishes to zero, which corresponds to a Gaussian distribution.
At the limit $\alpha \rightarrow 0$,  $\gamma_{np}=-\frac{2}{3}\left(\frac{1}{1+\frac{1}{\epsilon_A}+\frac{1}{\alpha\,\epsilon_A}}\right)^2$, which is always negative. Notice that, for finite stiffness the term $\alpha\,\epsilon_A$ must be non-zero even at $\alpha \rightarrow 0$ limit if the active noise is operating on, as this term determines its strength of correlation.  
 For any other values of $\alpha$, $\gamma_{np}$ takes negative value with a minimum occurring at $\alpha=\frac{1}{2}\frac{1+\sqrt{25+24 \epsilon_A}}{1+\epsilon_A}$, indicating the distribution decays faster than Gaussian. For any finite value of $\alpha$,  $\gamma_{np}$ is monotonically decreasing function with respect to $\epsilon_A$, starting with $\gamma_{np}=0$ at the limit $\epsilon_A \rightarrow 0$. It suggests, for small diffusivity of active noise the distribution becomes Gaussian centered at $\tilde{x}_f=0$ as shown in  Fig. \ref{plot_pst_epsilon_var}.

 To get an entire description of the dynamics, we need to perform the Fourier inversion as given in Eq. (\ref{pdf_dichosho_1}) which is difficult to do.
However, at the long time limit $\tilde{T} \rightarrow \infty $, the characteristic function can be simplified to  
\begin{equation}
\underset{T \rightarrow \infty}{\mathcal{L}t}G(\tilde{p}_T)=G_0(\tilde{p}_T)={}_{0}F_1\left(\frac{\alpha+1}{2},-\frac{\tilde{p}_T^2 \alpha^2}{4}\right).
\end{equation}
Therefore the distribution takes a simple form as
\begin{align}
\mathbb{P}_{st}(\tilde{x}_f)= \frac{1}{2 \pi}\int_{-\infty}^{+\infty}d\tilde{p}_T\,e^{-i \tilde{p}_T \tilde{x}_f}\,e^{-\frac{\alpha}{2 \epsilon_A}\tilde{p}_T^2}\,G_0(\tilde{p}_T)\label{pst_dicho}.
\end{align}
Along with  Gaussian Fourier transform $\frac{1}{2\pi}\int_{-\infty}^{+\infty}d\tilde{p}_T\,e^{-i \tilde{p}_T \tilde{x}_f}\,e^{-\frac{\alpha}{2 \epsilon_A}\tilde{p}_T^2}=\sqrt{\frac{\epsilon_A}{2 \pi \,\alpha}}\,e^{-\frac{\epsilon_A}{2  \,\alpha}\,\tilde{x}_f^2} $,  the following transform is required for calculation of $\mathbb{P}_{st}(x_f)$:
\begin{align}
\int_{-\infty}^{+\infty}d\tilde{p}_T\,e^{-i \tilde{p}_T \tilde{x}}\,G_0(\tilde{p}_T)=\frac{\Gamma(\frac{\alpha+1}{2})}{\sqrt{\pi}\Gamma(\frac{\alpha}{2})}\,\frac{1}{\alpha}\left(1-\frac{ \tilde{x}^2}{\alpha^2}\right)^{\alpha/2-1}\,\left(1-\Theta(1-\frac{\alpha^2}{ \tilde{x}^2})\right)
\label{fourier}.
\end{align}
Hence, the steady state probability can be rewritten as convolution of the above two Fourier transforms as
\begin{align}
\mathbb{P}_{st}(\tilde{x}_f)
=\frac{\Gamma(\frac{\alpha+1}{2})}{\sqrt{\pi}\Gamma(\frac{\alpha}{2})}\,\frac{1}{\alpha}\,\sqrt{\frac{\epsilon_A}{2\pi \alpha}}\,\int_{-\alpha}^{+\alpha}d\tilde{y}\,e^{-\frac{\epsilon_A}{2  \,\alpha} (\tilde{x}_f-\tilde{y})^2}\,\left(1-\frac{\tilde{y}^2}{\alpha^2}\right)^{\alpha/2-1}
\label{pst}.
\end{align}
We need to  compute  Eq. (\ref{pst}) which  is not possible analytically.  However one can consider the following limiting cases, for which the analysis can be carried out.
\subsubsection{ $\epsilon_A >>1$.}
 This limit corresponds to a situation where the  width of Gaussian, $\alpha/\epsilon_A$ is much smaller than $\alpha$ which is the  width of other function involved in the convolution with Gaussian in Eq. (\ref{pst}). Hence,
 \[\sqrt{\frac{\epsilon_A}{2 \pi \,\alpha}}\,e^{-\frac{\epsilon_A}{2  \,\alpha}} (\tilde{x}_f-\tilde{y})^2 \approx \delta(\tilde{x}_f-\tilde{y}).\] 
 Therefore, the steady state distribution becomes 
\begin{align}
\mathbb{P}_{st}(\tilde{x}_f)=\frac{\Gamma(\frac{\alpha+1}{2})}{\sqrt{\pi}\Gamma(\frac{\alpha}{2})}\,\frac{1}{\alpha}\,\left(1-\frac{\tilde{x}_f^2}{\alpha^2}\right)^{\alpha/2-1}\label{steadystate-bacteria}.
\end{align}
This result had been reported for non-interacting run-and-tumble particles confined within a circular pore \cite{RevModPhys.88.045006,0295-5075-86-6-60002}.  Here, the effect of thermal noise is negligible compared to active noise and as a consequence the distribution remains confined in the region [$-\alpha,+\alpha$].
\subsubsection{ $\epsilon_A \ll 1 $.}
In this case, normal diffusivity prevails over the  active one. Taking $\alpha\ll \frac{\alpha}{\epsilon_A}$ or $\epsilon_A  \ll 1 $, one can approximate
$ \frac{\Gamma(\frac{\alpha+1}{2})}{\sqrt{\pi}\Gamma(\frac{\alpha}{2})}\,\frac{1}{\alpha}\,\left(1-\frac{\tilde{x}_f^2}{\alpha^2}\right)^{\alpha/2-1} \approx \delta(\tilde{x}_{f}) $
to get 
\begin{align}
\mathbb{P}_{st}(\tilde{x}_f)&=
\sqrt{\frac{\epsilon_A}{2\pi \alpha}}\,e^{-\frac{\epsilon_A}{2  \,\alpha} \tilde{x}_f^2}.
\end{align}
Therefore for small value of $\epsilon_A$, distribution is Gaussian with a long spread, centered at $\tilde{x}_{f}=0$,  as pictorially depicted in  Fig. \ref{plot_pst_epsilon_var}.
\begin{figure}[h!]
\includegraphics[width=0.8\linewidth]{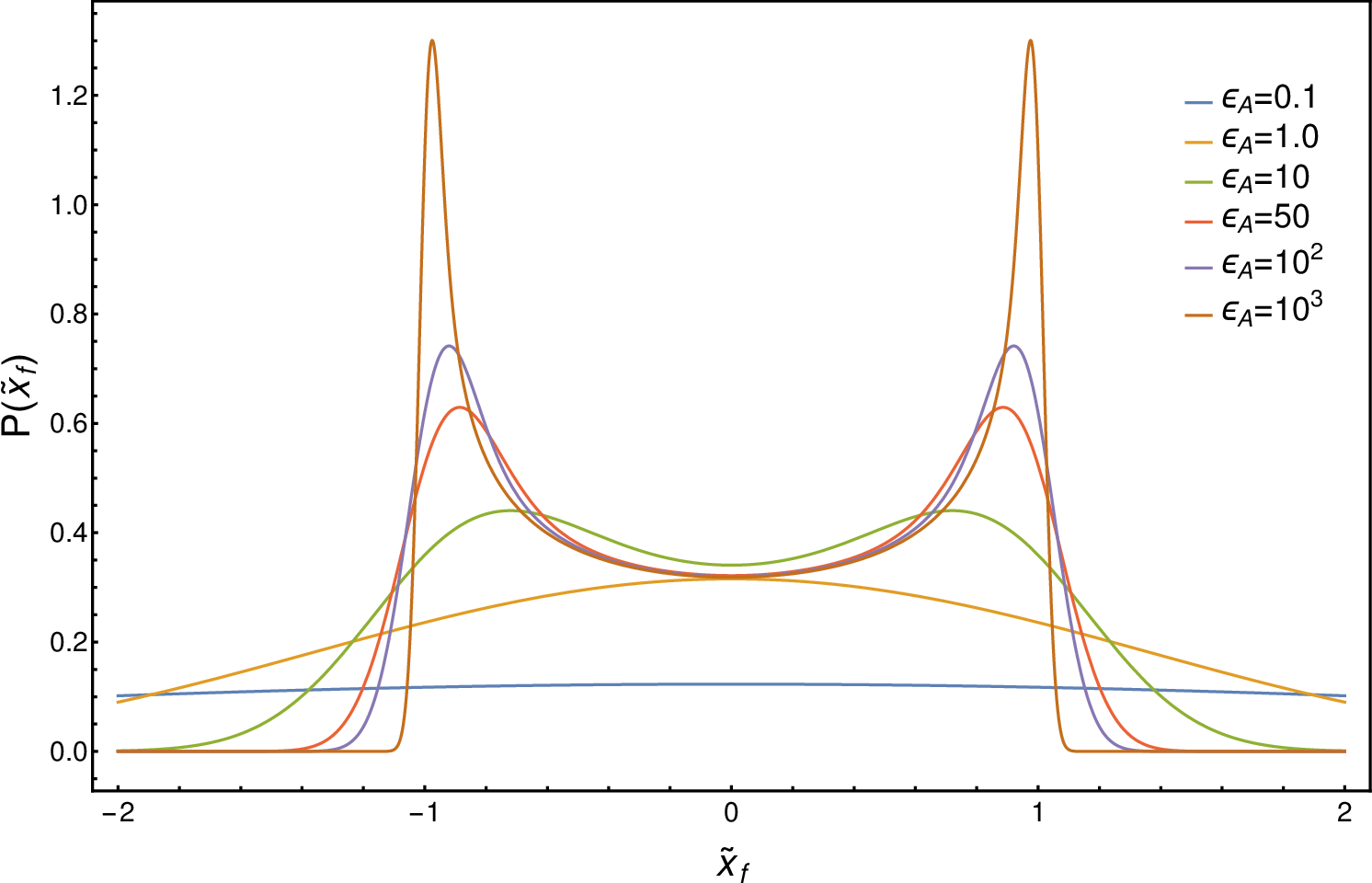}
\caption{Steady state distribution of particle confined in a harmonic potential in dichotomous noise is plotted against displacement for different values of $\epsilon_A$ at $\alpha=1.$}\label{plot_pst_epsilon_var}
\end{figure} 
\subsubsection{  $\alpha \gg 1$.}
The Fourier transform of $G_0(\tilde{p}_T)$ given in Eq. (\ref{fourier}) can be approximated at large $\alpha$ limit as given below.  

  According to Stirling's approximation, for large $\beta$,
	$\Gamma(\beta)\approx\sqrt{\frac{2\pi}{\beta}}\left(\frac{\beta}{e}\right)^{\beta}\,\left(1+\mathcal{O}(1/\beta)\right).$ 
	Using this,   for $\alpha\gg1$,one can approximate
	$ \frac{\Gamma(\frac{\alpha+1}{2})}{\Gamma(\frac{\alpha}{2})}\approx \sqrt{\frac{\alpha}{2}}
	$ and 	
 $\left(1+\frac{1}{\alpha}\right)^{\frac{\alpha}{2}}\approx\sqrt{e}.$
	Further,
$\left(1-\frac{ \tilde{y}^2}{\alpha^2}\right)^{\alpha/2-1} \simeq  e^{\text{ln}\left(1-\frac{ \tilde{y}^2}{\alpha^2}\right)^{\alpha/2}} \simeq  e^{-\frac{\tilde{y}^2}{2 \alpha}-\mathcal{O}\left(\frac{1}{\alpha^3}\right)} \sim e^{-\frac{\tilde{y}^2}{2 \alpha}}.
$
Therefore at very large $\alpha$ limit the probability distribution from Eq. (\ref{pst}) can be written as
\begin{align}
\mathbb{P}_{st}(\tilde{x}_f)
=\sqrt{\frac{1}{2\pi\alpha}}\,\sqrt{\frac{\epsilon_A}{2\pi\alpha}}\,\int_{-\infty}^{+\infty}d\tilde{y}\,e^{-\frac{\epsilon_A}{2  \,\alpha} (\tilde{x}_f-\tilde{y})^2}\,e^{-\frac{\tilde{y}^2}{2 \alpha}}=\sqrt{\frac{1}{2\pi\alpha\left(1+\frac{1}{\epsilon_A}\right)}}\,e^{-\frac{\tilde{x}_f^2}{2\pi\alpha\left(1+\frac{1}{\epsilon_A}\right)}}\label{ana_pst_highc}.
\end{align}
For finite stiffness of potential, large $\alpha$ regime reflects large $\gamma$, which becomes delta correlated at this limit.  As a consequence the steady state distribution is Gaussian as given in Eq. (\ref{ana_pst_highc}).
\subsubsection{$\alpha \ll 1$.}
 At very small $\alpha$ limit,  one can write:  $\frac{\Gamma(\frac{\alpha+1}{2})}{\Gamma(\frac{\alpha}{2})} \approx \frac{\alpha \sqrt{\pi} }{2}.
$
The other term in $G_0(\tilde{p}_T)$ can be  approximated as
\begin{align}
\left(1-\frac{ \tilde{y}^2}{\alpha^2}\right)^{\alpha/2-1} \approx e^{-\text{ln}\left(1-\frac{ \tilde{y}^2}{\alpha^2}\right) }\approx \delta\left(\tilde{y}-\alpha\right)+\delta\left(\tilde{y}+\alpha\right) 
\label{c0fourier}.
\end{align}
Therefore by virtue of Eq. (\ref{fourier}) and Eq. (\ref{c0fourier}), Eq. (\ref{pst}) can be computed as
\begin{align}
\mathbb{P}_{st}(\tilde{x}_f)
=\frac{1}{2}\,\sqrt{\frac{\epsilon_A}{2\pi\alpha}}\left[\,e^{-\frac{\epsilon_A}{2  \,\alpha} (\tilde{x}_f-\alpha)^2}+\,e^{-\frac{\epsilon_A}{2  \,\alpha} (\tilde{x}_f+\alpha)^2}\right] \approx \frac{1}{2}\delta(\tilde{x}_f-\alpha)+\frac{1}{2}\delta(\tilde{x}_f+\alpha)\label{ana_pst_smallc}.
\end{align}
Small $\alpha$ implies the very slow Poisson rate. Hence one should expect particle to be gathered almost deterministically at two positions $+\alpha,\,-\alpha$. The analytical approximate result given in Eq. (\ref{ana_pst_smallc}) as well as the plot  in Fig. \ref{plot_pdf-dicho_alpha}  support this statement.
\begin{figure}[H]
	\includegraphics[width=0.8\linewidth]{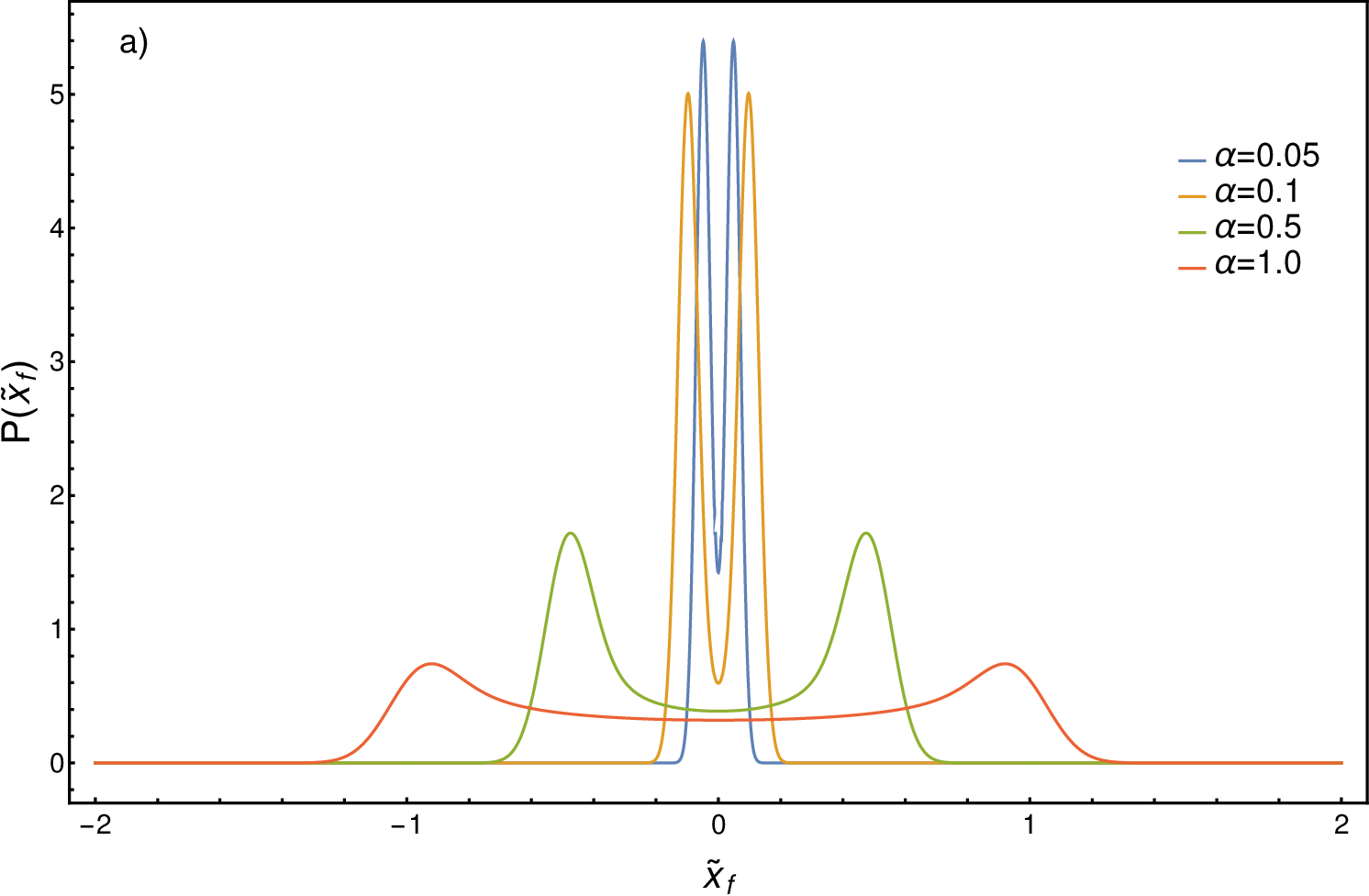} \quad 
	\includegraphics[width=0.8\linewidth]{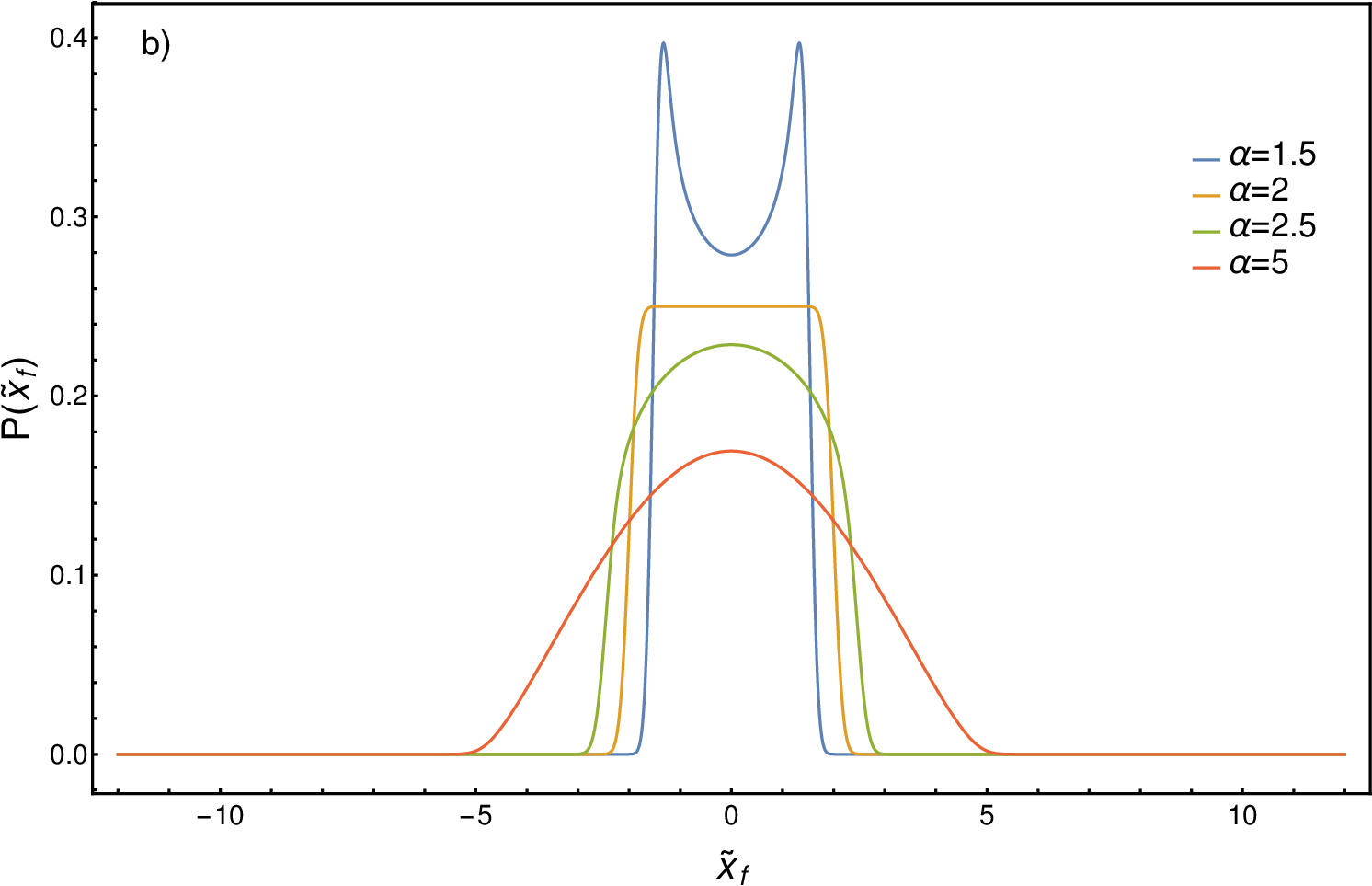}
	\caption{Steady state distribution of a confined particle in presence of dichotomous noise is plotted against scaled displacement for different values of $\alpha$ at $\epsilon_A=100$.} \label{plot_pdf-dicho_alpha}
\end{figure}
\subsubsection{$\alpha = 2$.}
 For $\alpha=2$, the integration involving in Eq. (\ref{pst}) can be evaluated exactly which results
\begin{align}
\mathbb{P}_{st}(\tilde{x}_f)=\frac{1}{4\alpha}\left[\text{erf}\left(\sqrt{\frac{\epsilon_A}{2\alpha}}[\tilde{x}_f+\alpha]\right)-\text{erf}\left(\sqrt{\frac{\epsilon_A}{2 \alpha}}[\tilde{x}_f-\alpha]\right)\right].
\end{align}
Notice that the distribution at $\alpha=2$ is almost flat, whereas for $\alpha<2$ the distribution is bimodal and for $\alpha>2 $ it is single peaked as illustrated in Fig. \ref{plot_pdf-dicho_alpha}. Therefore $\alpha=2$ represents a crossover regime. It is interesting to note here that similar trait has been observed for a model with switching environments where alteration of switching rate from slow to fast transits the distribution from bimodal to unimodal \cite{hufton2016intrinsic}.
	\subsection{ Poissonian White Noise}
Here we consider  a particle undergoing random walk in a harmonic potential in presence of both Gaussian white noise $\eta(t)$
and Poissonian white noise $\sigma_{PW}(t)$. The particle is initially
Gaussian distributed with a distribution: $P(x_{0})=\sqrt{\frac{\lambda}{2\pi D}}\,e^{-\frac{\lambda}{2D}\,x_{0}^{2}}$.  
 Using Eq. (\ref{eq:ch functional Poisson}) we have calculated characteristic
functional which is  given as 
\begin{equation}
\mathbb{G}(p_{T})=\left[\frac{1+a_{0}^{2}p_{T}^{2}e^{-2\lambda T}}{1+a_{0}^{2}p_{T}^{2}}\right]^{\frac{\mu}{2\lambda}}\label{Fsho}.
\end{equation}
Therefore from Eq. (\ref{eq:finP3sho}) one can express the propagator as
\begin{equation}
\mathbb{P}(x_{f},T;\,x_{0},0)=\frac{1}{2\pi}\int_{-\infty}^{+\infty}dp_{T}\,e^{-ip_{T}x_{f}+ip_{0}x_{0}-\frac{D\,p_{T}^{2}}{2\lambda}(1-e^{-2\lambda T})}\,\mathbb{G}(p_{T})\label{Pshopsn}.
\end{equation}
The relaxation time of the particle in the harmonic potential is denoted here
as $\tau_{r}$ and is given by: $\tau_{r}=\frac{1}{\lambda}$. The
characteristic timescale to occur a Poisson pulse is defined as: $\tau_{a}=\frac{1}{\mu}$. So we can write down a dimensionless parameter $\tilde{\tau}$ as $\tilde{\tau}=\frac{\tau_a}{\tau_r}$. Using Eq. (\ref{Pshopsn}) into Eq. (\ref{Probfinal}) one can obtain the probability
of finding the particle at position $x_{f}$ at time $T$ and it reads
\begin{equation}
\mathbb{P}(x_{f},T)=\frac{1}{2\pi}\int_{-\infty}^{+\infty}dp_{T}\,e^{-ip_{T}x_{f}}\,\mathbb{F}(p_{T})\label{Probfinsho}.
\end{equation}
Here, 
\begin{align}
 \mathbb{F}(p_{T})&=\sqrt{\frac{\lambda}{2\pi D}}\,\int_{-\infty}^{+\infty}dx_{0}\,e^{ip_{T}e^{-\lambda T}\,x_{0}-\frac{D\,p_{T}^{2}}{2\lambda}(1-e^{-2\lambda T})}\,\mathbb{G}(p_{T})\,e^{-\frac{\lambda}{2D}\,x_{0}^{2}}\nonumber\\
 &=e^{-\frac{D}{2\lambda}\,p_{T}^{2}}\,\left[\frac{1+a_{0}^{2}p_{T}^{2}e^{-2\lambda T}}{1+a_{0}^{2}p_{T}^{2}}\right]^{\frac{1}{2\tilde{\tau}}}\label{Fshopwn}.
\end{align}

Using Eq. (\ref{Fsho}) in Eq. (\ref{m-th moment}) we have computed
second and fourth moment and obtain non-Gaussian parameter by virtue
of relation (\ref{NGP}). The MSD is given as
\begin{align}
<x^{2}(T)>=\frac{D}{\lambda}+\frac{D_{A}}{\lambda}(1-e^{-2\lambda T}),\label{msd_pwn_sho}
\end{align}
and the fourth moment can be expressed as
\[\left<x^4(T)\right>=3 \left<x^2(T)\right>^2+6\,D_A \,\frac{a^2}{\lambda}\,(1-e^{-4T\lambda}).\]
For the time limit $T<\tau_{r}$, dynamics is diffusive as
 \[
<x^{2}(T)>=2\,D_{A}\,T+D\,\tau_{r}.
\]
For very large $T$ limit, 
\[
<x^{2}(T)>=(D+D_{A})\,\tau_{r}.
\]
MSD is independent of time at very large time suggesting that the
system reaches to a steady state. The NGP, $\gamma_{np}$ is determined using Eq. (\ref{NGP}) and it is given as
\begin{align}
\gamma_{np}&=\frac{6\,D_A \,\frac{a^2}{\lambda}\,(1-e^{-4T\lambda})}{3(\frac{D}{\lambda}+\frac{D_{A}}{\lambda}(1-e^{-2\lambda T}))^2}\nonumber\\
&=\frac{4\,\tilde{\tau}\,D_{A}^{2}\,\sinh(2\lambda T)}{((D+2D_{A})\sinh(\lambda T)+D\cosh(\lambda T))^{2}}\nonumber\\
&=\frac{4\,\tilde{\tau}\,\epsilon_A^{2}\,\sinh(2\tilde{T})}{\left((1+2\epsilon_{A})\sinh(\tilde{T})+\cosh(\tilde{T})\right)^{2}}.
\end{align}
Here we have described the ratio of two diffusivities by a dimensionless parameter $\epsilon_A$ which is given as: $\epsilon_{A}=\frac{D_{A}}{D}$ and defining a dimensionless parameter $\tilde{T}$ as $\tilde{T}=\lambda\,T.$ At small time limit $\gamma_{np}$ vanishes suggesting Gaussian behavior of PDF. 
\begin{figure}[h!]
 \includegraphics{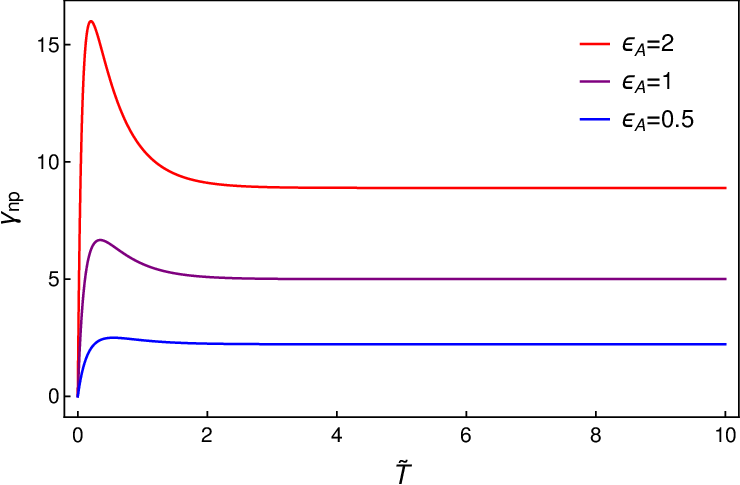} 
\caption{ Non-Gaussian parameter $(\gamma_{np})$ is plotted as a function of scaled time $ \tilde{T}$ for three different values of $\epsilon_{A}$
at $\tilde{\tau}=10$.
The feature of distribution curve remains unchanged for any values of $\tilde{\tau}$.}
\label{ngpshopwnplot} 
\end{figure}

From Fig. \ref{ngpshopwnplot} one can observe that $\gamma_{np}$
starts from zero and initially increases with time. At an intermediate
time it attains a peak indicating strong non-Gaussian characteristics.
Then it keep decreasing and converges to a finite positive non-zero
constant value of $\frac{2\,\tilde{\tau}\,\epsilon_{A}^{2}}{(1+\epsilon_{A})^{2}}.$ It implies that the distribution attains a steady state  which is non-Gaussian having slow-decaying tail compared to Gaussian.

The full description of the dynamics can be investigated by computing probability distribution function (PDF). At this stage we have made the propagator dimensionless by defining some dimensionless variables as: $\tilde{T}=\frac{T}{\tau_r},\,\tilde{x}_f=\frac{x_f}{a_0},\,\tilde{p_T}=p_T\,a_0$.  So the propagator becomes
\begin{align}
&\mathbb{P}(\tilde{x}_{f},\tilde{T})=\frac{1}{2\pi}\int_{-\infty}^{+\infty}d\tilde{p}_{T}\,e^{-i\tilde{p}_{T}\,\tilde{x}_{f}}\,\mathbb{F}(\tilde{p}_{T})\label{Probfinsho_dless},
\end{align}
where \begin{align}
&\mathbb{F}(\tilde{p}_{T})=e^{-\frac{\tilde{p}_{T}^{2}}{2\epsilon_A\,\tilde{\tau}}}\,\left[\frac{1+\tilde{p}_{T}^{2}e^{-2\tilde{ T}}}{1+\tilde{p}_{T}^{2}}\right]^{\frac{1}{2\tilde{\tau}}}
\label{Probfinsho_dless_1}.
\end{align}
From Eq. (\ref{Probfinsho_dless}) it is clear that  the PDF cannot be calculated
analytically. However one can consider different time scales as discussed below.
\subsubsection{Short time  ($\tilde{T}\ll 1 $).}
 At small time and position region $i.e.$ at the limit $\tilde{p}_{T}^{2}\gg1$ and $\tilde{T} \ll 1 $, one can approximate the characteristic
functional $\mathbb{F}(\tilde{p}_{T})$  in Eq. (\ref{Probfinsho_dless_1}) as
\begin{align}
\mathbb{F}(\tilde{p}_{T})\sim e^{-\frac{\tilde{p}_{T}^{2}}{2\epsilon_A\,\tilde{\tau}}-\frac{\tilde{T}}{\tilde{\tau}}},
\end{align}
which gives
\begin{equation}
\mathbb{P}(\tilde{x}_{f},\tilde{T})\sim \sqrt{\frac{\epsilon_A\,\tilde{\tau}}{2 \pi }} e^{-\frac{\epsilon_A\,\tilde{\tau}}{2}\,\tilde{x}_{f}^{2}-\frac{\tilde{T}}{\tilde{\tau}}}\label{pdf_sho_smallxt}.
\end{equation}
Therefore the distribution is Gaussian at small $\tilde{x}_f$ whose strength is decaying in time due to the effect of active noise.
\subsubsection{Intermediate time.} 
  At small displacement region, $i.e.,$ at $\tilde{x}_f\ll 1$, $\tilde{p}_{T}\gg1$  and
 $\mathbb{F}(\tilde{p}_{T})\sim e^{-\frac{\tilde{p}_{T}^{2}}{2\epsilon_A\,\tilde{\tau}}-\frac{\tilde{T}}{\tilde{\tau}}}.$
 Therefore the distribution near the origin is
 \begin{equation}
 \mathbb{P}(\tilde{x}_{f},\tilde{T})\sim \sqrt{\frac{\epsilon_A\,\tilde{\tau}}{2 \pi }} e^{-\frac{\epsilon_A\,\tilde{\tau}}{2}\,\tilde{x}_{f}^{2}-\frac{\tilde{T}}{\tilde{\tau}}}.
 \end{equation}
In order to get the distribution for other values of $\tilde{x}_{f}$,  one write \begin{align}
\left[\frac{1+\tilde{p}_{T}^{2}e^{-2\tilde{T}}}{1+\tilde{p}_{T}^{2}}\right]^{\frac{1}{2\tilde{\tau}}}=\left[e^{-2\tilde{T}}+\frac{1-e^{-2\tilde{T}}}{1+\tilde{p}_{T}^{2}}\right]^{\frac{1}{2\tilde{\tau}}},
\end{align}
and expand as the following binomial series:
 \begin{equation}
\left[e^{-2\tilde{T}}+\frac{1-e^{-2\tilde{T}}}{1+\tilde{p}_{T}^{2}}\right]^{\frac{1}{2\tilde{\tau}}}=\sum_{n=0}^{n=\infty}\binom{\frac{1}{2\tilde{\tau}}}{n}\left(e^{-2\tilde{T}}\right)^{\frac{1}{2\tilde{\tau}}-n}\frac{\left(1-e^{-2\tilde{T}}\right)^n}{\left(1+\tilde{p}_{T}^{2}\right)^n}. \label{expan}
 \end{equation}
  Using the  convolution theorem  (Eq. (\ref{G})) with the help of Eq. (\ref{fouriertransformCauchy})-(\ref{fouriertransformGaussian}),
  one can rewrite the propagator in Eq. (\ref{Probfinsho_dless}) as
  \begin{align}
  \mathbb{P}&(\tilde{x}_{f},\tilde{T})=\sqrt{\frac{\epsilon_{A}\tilde{\tau}}{2\pi}}\,e^{-\frac{\tilde{T}}{\tilde{\tau}}-\frac{\epsilon_{A}\,\tilde{\tau}}{2}\,\widetilde{x}_{f}^{2}}\nonumber \\
  & +\frac{1}{2\pi}\sum_{n=1}^{\infty}\binom{\frac{1}{2\tilde{\tau}}}{n}e^{-\frac{\tilde{T}}{\tilde{\tau}}}\left(e^{2\tilde{T}}-1\right)^n\,\int_{-\infty}^{\infty}d\tilde{x}\frac{2^{\frac{3}{2}-n}\,\sqrt{\pi}\,\left|\tilde{x}\right|^{n-\frac{1}{2}}K_{n-\frac{1}{2}}\left(|\tilde{x}|\right)}{\Gamma(n)}\sqrt{\frac{\epsilon_{A}\tilde{\tau}}{2\pi}}e^{-\epsilon_{A}\tilde{\tau}(\tilde{x_{f}}-\tilde{x})^{2}/2}\label{p(x)pwnsmallx22}.
  \end{align}
  Now considering if the Gaussian has a much narrower width than the Bessel function, $i.e.$,   $\epsilon_A\,\tilde{\tau}\gg 1,$ one can approximate  the Gaussian as a delta function, and then  from Eq. (\ref{p(x)pwnsmallx22})  one has
  \begin{align}
  \mathbb{P}&(\tilde{x}_{f},\tilde{T})=\sqrt{\frac{\epsilon_{A}\tilde{\tau}}{2\pi}}\,e^{-\frac{\tilde{T}}{\tilde{\tau}}-\frac{\epsilon_{A}\,\tilde{\tau}}{2}\,\widetilde{x}_{f}^{2}}\nonumber \\
  & +\frac{1}{2\pi}\sum_{n=1}^{\infty}\binom{\frac{1}{2\tilde{\tau}}}{n}e^{-\frac{\tilde{T}}{\tilde{\tau}}}\left(e^{2\tilde{T}}-1\right)^n\,\frac{2^{\frac{3}{2}-n}\,\sqrt{\pi}\,\left|\tilde{x}_f\right|^{n-\frac{1}{2}}K_{n-\frac{1}{2}}\left(|\tilde{x}_f|\right)}{\Gamma(n)}.\label{p(x)pwnsmallx21}
  \end{align}
   Strictly speaking the  expansion of Eq. (\ref{expan}) requires $e^{-2\tilde{T}}>(1-e^{-2\tilde{T}})$ or $\tilde{T}<\ln(2)/2$. However, the expansion of Eq. (\ref{p(x)pwnsmallx21}) seems to be valid for other values of $\tilde{T}$ too.
  
  To get the tail behavior of the distribution, one can use the asymptotic form for the Bessel function $K_{n}\left(|\tilde{x}_{f}|\right)\approx\sqrt{\frac{\pi}{2\mid \tilde{x}_{f}\mid}}\,e^{-\mid \tilde{x}_{f}\mid}$ in the above, and thereafter summing over  $n$ one can obtain 
  \begin{equation}
  \mathbb{P}(\tilde{x}_{f},\tilde{T};\,\tilde{x}_{0}=0,0)=\sqrt{\frac{\epsilon_A\,\tilde{\tau}}{2 \pi }} e^{-\frac{\epsilon_A\,\tilde{\tau}}{2}\,\tilde{x}_{f}^{2}-\frac{\tilde{T}}{\tilde{\tau}}}+\frac{e^{-|\tilde{x}_{f}|-\frac{\tilde{T}}{\tilde{\tau}}}}{4\tilde{\tau}}\left(e^{2\tilde{T}}-1\right){}_1F_1\left(1-\frac{1}{2\tilde{\tau}};2;-\frac{1}{2}(e^{2\tilde{T}}-1)\mid \tilde{x}_{f}\mid\right)\label{p(x)pwnshosmallfin}.
  \end{equation}
So the distribution has a Gaussian core near the  origin as given by the first term on the right-hand side of Eq. (\ref{p(x)pwnshosmallfin}). The second term signifies the departure from  Gaussianity, which decays like \[\mathbb{P}(\tilde{x}_{f},\tilde{T}) \sim \frac{2^{1-\frac{1}{2\tilde{\tau}}}\left(e^{2\tilde{T}}-1\right)^{\frac{1}{2\tilde{\tau}}}}{\Gamma\left(1+\frac{1}{2\tilde{\tau}}\right)|\tilde{x}_f|^{1-\frac{1}{2\tilde{\tau}}}}\, \frac{e^{-|\tilde{x}_{f}|-\frac{\tilde{T}}{\tilde{\tau}}}}{4\tilde{\tau}}.\] This analytical result  is in agreement with numerical calculations, as shown in  Fig. \ref{pdf_sho_smallt}.
 
 To capture the sensitivity of initial distribution on dynamics at transient period, one can  consider the particle's initial distribution  as a delta function (i.e. $\delta(x_0)$) and hence,   Eq. (\ref{Probfinsho}) can be reconstructed as
	\begin{equation}
	\mathbb{P}(x_{f},T)=\frac{1}{2\pi}\int_{-\infty}^{+\infty}dp_{T}\,e^{-ip_{T}x_{f}}\,\mathbb{F}(p_{T})\label{P_transient1},
	\end{equation} 
	with 
	\begin{align}
	\mathbb{F}(p_{T})=e^{-\frac{D\,p_{T}^{2}}{2\lambda}(1-e^{-2\lambda T})}\,\left[\frac{1+a_{0}^{2}p_{T}^{2}e^{-2\lambda T}}{1+a_{0}^{2}p_{T}^{2}}\right]^{\frac{1}{2\tilde{\tau}}}\label{F_transient1}.
	\end{align}  
Upon rescaling as earlier, the above can be expressed as
\begin{align}
\mathbb{P}(\tilde{x}_{f},\tilde{T})=\frac{1}{2\pi}\int_{-\infty}^{+\infty}d\tilde{p}_{T}\,e^{-i\tilde{p}_{T}\,\tilde{x}_{f}}\,e^{-\frac{\tilde{p}_{T}^{2}}{2\epsilon_A\,\tilde{\tau}}(1-e^{-2 \tilde{ T}})}\,\left[\frac{1+\tilde{p}_{T}^{2}e^{-2\tilde{ T}}}{1+\tilde{p}_{T}^{2}}\right]^{\frac{1}{2\tilde{\tau}}}\label{P_trans1},
\end{align}
which can be rewritten as 
\begin{align}
\mathbb{P}&(\tilde{x}_{f},\tilde{T})=\sqrt{\frac{\epsilon_{A}\tilde{\tau}}{2\left(1-e^{-2\tilde{T}}\right)\pi}}\,e^{-\frac{\tilde{T}}{\tilde{\tau}}-\frac{\epsilon_{A}\,\tilde{\tau}}{2\left(1-e^{-2\tilde{T}}\right)}\,\widetilde{x}_{f}^{2}}\nonumber \\
& +\frac{1}{2\pi}\sum_{n=1}^{\infty}\binom{\frac{1}{2\tilde{\tau}}}{n}e^{-\frac{\tilde{T}}{\tilde{\tau}}}\left(e^{2\tilde{T}}-1\right)^n\,\int_{-\infty}^{\infty}d\tilde{x}\frac{2^{\frac{3}{2}-n}\,\sqrt{\pi}\,\left|\tilde{x}\right|^{n-\frac{1}{2}}K_{n-\frac{1}{2}}\left(|\tilde{x}|\right)}{\Gamma(n)}\sqrt{\frac{\epsilon_{A}\tilde{\tau}}{2\pi}}e^{-\frac{\epsilon_{A}\tilde{\tau}}{2\left(1-e^{-2\tilde{T}}\right)}(\tilde{x_{f}}-\tilde{x})^{2}}\label{P_transient2}.
\end{align}

Therefore similar to the previous analysis, in the limit $\frac{\epsilon_{A}\tilde{\tau}}{\tilde{T}}\gg 1,$ the distribution can be given as
\begin{align}
\mathbb{P}(\tilde{x}_{f},\tilde{T};\,\tilde{x}_{0}=0,0)&\approx \sqrt{\frac{\epsilon_{A}\tilde{\tau}}{2\left(1-e^{-2\tilde{T}}\right)\pi}}\,e^{-\frac{\tilde{T}}{\tilde{\tau}}-\frac{\epsilon_{A}\,\tilde{\tau}}{2\left(1-e^{-2\tilde{T}}\right)}\,\widetilde{x}_{f}^{2}}\nonumber\\
&+\frac{e^{-|\tilde{x}_{f}|-\frac{\tilde{T}}{\tilde{\tau}}}}{4\tilde{\tau}}\left(e^{2\tilde{T}}-1\right){}_1F_1\left(1-\frac{1}{2\tilde{\tau}};2;-\frac{1}{2}(e^{2\tilde{T}}-1)\mid \tilde{x}_{f}\mid\right)
\label{p(x)_transient}.
\end{align}
So, the choice of different initial conditions only modifies the central core  but the tail behavior remains unchanged.
\subsubsection{Long time, $\tilde{T}\gg 1$.}
At long time limit,  Eq. (\ref{Probfinsho_dless_1}) can be simplified by taking  $e^{-2\tilde{ T}}\rightarrow 0,$ and  approximating $\mathbb{F}(p_{T})$ as 
\begin{equation}
\mathbb{F}(\tilde{p}_{T})\approx e^{-\frac{\tilde{p}_{T}^{2}}{2\epsilon_A\,\tilde{\tau}}}\,\left(\frac{1}{1+\tilde{p}_{T}^{2}}\right)^{\frac{1}{2\tilde{\tau}}}\label{Flargex}.\end{equation}
Hence, the distribution becomes
\begin{align}
\mathbb{P}(\tilde{x}_{f},\tilde{T})=\frac{1}{2\pi}\int_{-\infty}^{+\infty}d\tilde{p}_{T}\,e^{-i\tilde{p}_{T}\,\tilde{x}_{f}}\,e^{-\frac{\tilde{p}_{T}^{2}}{2\epsilon_A\,\tilde{\tau}}}\,\left(\frac{1}{1+\tilde{p}_{T}^{2}}\right)^{\frac{1}{2\tilde{\tau}}}. \label{Probfinsho_dless_rewritting}
\end{align}

$3.1.\; \tilde{x}_f\ll 1$

In small $\tilde{x}_f$ limit, $\tilde{p}_{T}^{2}\gg1$. So, it allows us to approximate $\mathbb{F}(\tilde{p}_{T})$  as 
\begin{equation}
\mathbb{F}(\tilde{p}_{T})\approx e^{-\frac{\tilde{p}_{T}^{2}}{2\epsilon_A\,\tilde{\tau}}}\:[\tilde{p}_{T}^{2}]^{-\frac{1}{2\tilde{\tau}}}\label{approx F}.
\end{equation}
After performing the inverse Fourier transform of relation (\ref{approx F}) with the assumption $\tilde{\tau}>1,$
one can obtain the PDF and it reads 
\begin{equation}
\mathbb{P}(\tilde{x}_{f})=\frac{\Gamma(\frac{1}{2}-\frac{1}{2\tilde{\tau}})}{2\,\pi}\,\left(\frac{1}{2\epsilon_A\,\tilde{\tau}}\right)^{-\frac{1}{2}(1-1/\tilde{\tau})}\,{}_{1}F_{1}\left(\frac{1}{2}-\frac{1}{2\tilde{\tau}};\frac{1}{2};-\epsilon_A\frac{\tilde{\tau}\,\tilde{x}_{f}^{2}}{2}\right)\label{PDF 1F1}.
\end{equation}
We have encountered the same kind of distribution in the section \ref{PSNsection}
and have seen that near to the origin, distribution is Gaussian of the form: 
\begin{equation}
\mathbb{P}(\tilde{x}_{f})\sim e^{-\frac{\epsilon_A\,\tilde{\tau}}{2}\,\tilde{x}_{f}^{2}}.
\end{equation}

 $3.2.\;\text{Intermediate }\,\tilde{x}_f$
 
To get the tail behavior, Eq. (\ref{Probfinsho_dless_rewritting}) is required to be evaluated which unfortunately cannot be done directly. However it can be rewritten as convolution of two functions in Fourier domain as discussed in section \ref{PWNsection} as well as in section \ref{PSNsection}.
So,  PDF in Eq. (\ref{Probfinsho_dless}) can be expressed as
\begin{align}
\mathbb{P}(\tilde{x}_f)=\int_{-\infty}^{+\infty}\,d\tilde{x}\,\sqrt{\frac{\epsilon_{A}\tilde{\tau}}{2\pi}}\,e^{-\frac{\epsilon_{A}\,\tilde{\tau}}{2}\,(\widetilde{x}_{f}-\tilde{x})^{2}}\times \frac{2^{\frac{3}{2}-\frac{1}{2\tilde{\tau}}}\,\sqrt{\pi}\,\left|\widetilde{x}\right|^{\frac{1}{2\tilde{\tau}}-\frac{1}{2}}K_{\frac{1}{2\tilde{\tau}}-\frac{1}{2}}\left(\left|\widetilde{x}\right|\right)}{\Gamma(\frac{1}{2\tilde{\tau}})}.
\end{align} 
For $\epsilon_A\,\tilde{\tau}\gg 1$, one can replace Gaussian distribution with a delta function as
\[\sqrt{\frac{\epsilon_A\,\tilde{\tau}}{2 \pi }} e^{-\frac{\epsilon_A\,\tilde{\tau}}{2}\,(\tilde{x}-\tilde{y})^2} \approx \delta(\tilde{x}-\tilde{y}).\]
Therefore, the approximate distribution is given by
\begin{equation}
\mathbb{P}(\tilde{x}_{f}) \approx\frac{2^{\frac{1}{2}-\frac{1}{2 \tilde{\tau}}}\,\left|\tilde{x}_{f}\right|^{\frac{1}{2\tilde{\tau}}-\frac{1}{2}}K_{\frac{1}{2 \tilde{\tau}}-\frac{1}{2}}\left(\left|\tilde{x}_{f}\right|\right)}{\sqrt{\pi}\,\Gamma(\frac{1}{2\tilde{\tau}})}\label{PDF bessel}.
\end{equation}
So, the distribution has an exponential
tail of the form : $e^{-\vert \tilde{x}_{f}\vert}$. In fact at $\tilde{T}\rightarrow\infty$ limit and for $2\tilde{\tau}=1$ one can compute the
distribution exactly as discussed in Appendix  \ref{appen 1} and
the probability is given in Eq. (\ref{P for Nu 1 }). It is conspicuous that the probability decays exponentially at large displacement and this is true for any values of $\tilde{\tau}$. However, the distribution for $2 \tilde{\tau}=1$  does not contain any attributes of Eq. (\ref{PDF 1F1}).  
As we are interested to know the steady state distribution we have
considered $ \tilde{T}\rightarrow\infty$ limit for different values of $\tilde{\tau}$
and have performed numerical integration of relation (\ref{eq:P s-integrand})
and get PDF vs. displacement plot as shown in Fig. \ref{pdfplotsho}. The plots are in accord with our analytical findings.
\\

\begin{figure}
\includegraphics[width=0.8\linewidth]{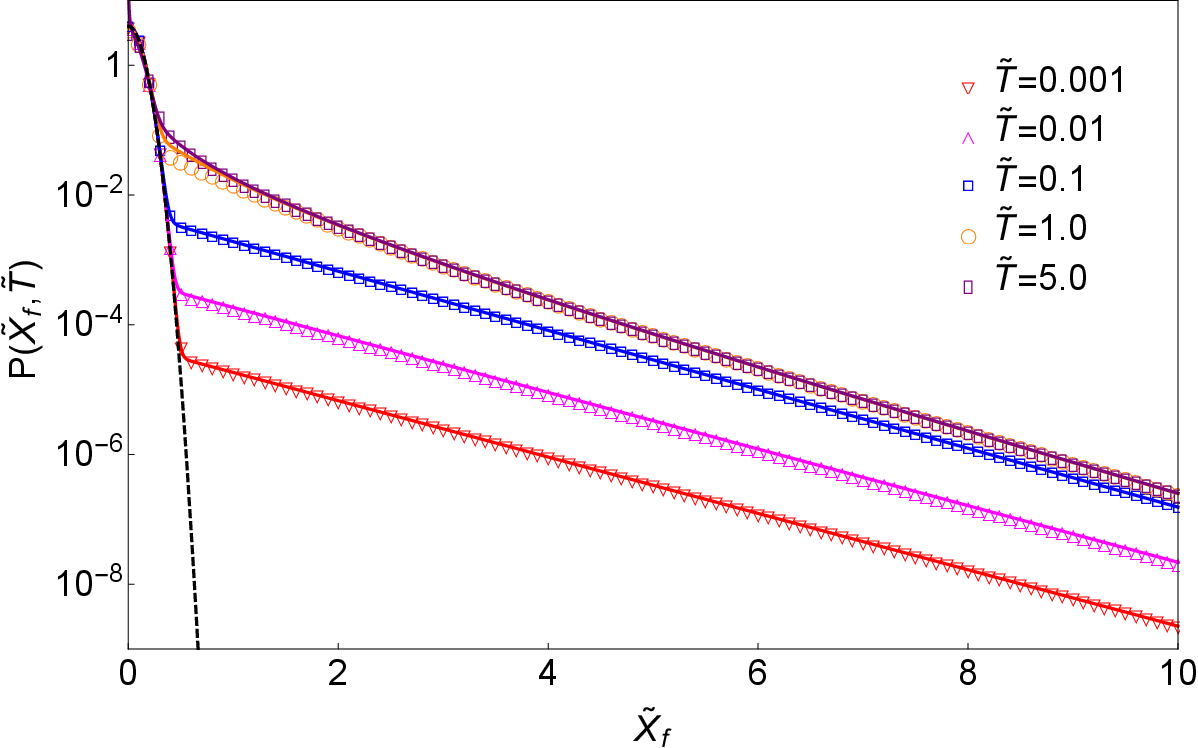} 
\caption{Logarithm of the probability distribution(PDF) function plotted against	$\tilde{x}_{f}$ at different timescales as presented with symbols. PDF is computed numerically for a particular set of parameters $\{\epsilon_A=10,\, \tilde{\tau}=10\}$ using Eq. (\ref{Probfinsho_dless}). The solid curves are obtained using Eq. (\ref{p(x)pwnshosmallfin}) which fit very well with the numerical curves.  The black dotted curve is the central	Gaussian part of the distribution   \ref{p(x)pwnshosmallfin}.}\label{pdf_sho_smallt} 
\end{figure}

\begin{figure}
\includegraphics[width=1\linewidth]{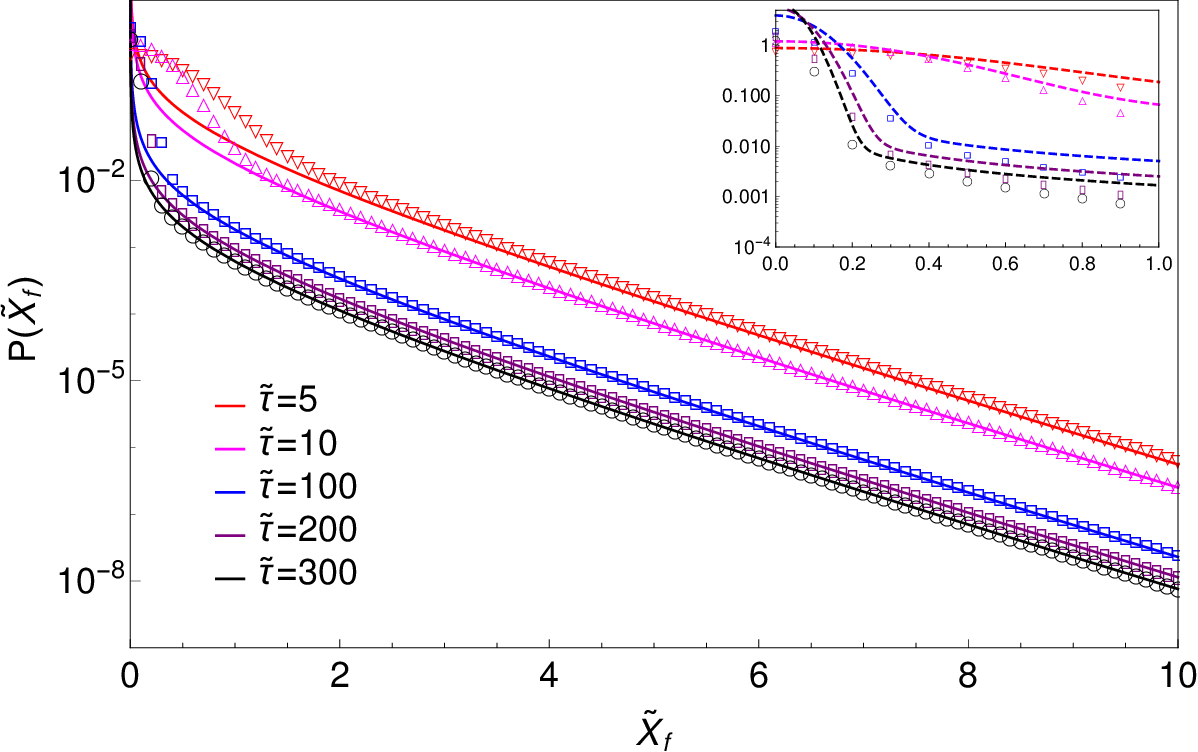} 
\caption{Logarithm of the probability distribution(PDF) function  plotted against 	displacement for
 different values of $\tilde{\tau}$ at $\epsilon_A=1$. 	PDF is computed numerically using Eq. (\ref{eq:P s-integrand}) and compare with Eq. (\ref{PDF bessel}) (solid lines) which are well fitted at intermediate to long displacement limit. At short distance, those agree well with Eq. (\ref{PDF 1F1})(dashed curves) as shown in inset.}
\label{pdfplotsho} 
\end{figure}

\section{Conclusion }\label{conclusion}

In the first part of the paper, we studied the diffusion of a free particle subject to both white and active noise. As expected, we have seen that in the case where both thermal and active noises are Gaussian, the PDF for the displacement of the particle is always Gaussian.   A system driven by active dichotomous noise, too gives a Gaussian distribution at long time scales. In the case where  active noise is dominant over the thermal noise, the approach to one single Gaussian distribution at long times is through the spreading of two Gaussians peaked at $+u\,T$ and $-u\,T$, analogous to what was seen in \cite{PhysRevE.88.032304}.   In the case of active noise being white Poissonian, we find that the distribution is Gaussian at very large time and displacement
limits as a consequence of the central limit theorem. But the initial
distribution is non-Gaussian, and has a central Gaussian part with an exponential tail.   This kind of distribution has been observed in some experiments \cite{PhysRevLett.103.198103,fodor2015activity}.  The presence of an exponential tail is a manifestation of Poisson noise, the amplitude of which is itself exponentially distributed. The dynamics in case of non-white (correlated) Poisson noise (results of Section \ref{PSNsection} ) has been found to be fascinating. In this
case the distribution for large displacements is  Gaussian at long time. But in an intermediate time scale, the distribution largely deviates from Gaussianity, having a central Gaussian part with prominent exponential tail.  This  has also been supported  by NGP calculation where NGP is found as a non-monotonic function of time and at intermediate time it attains a maximum.  These results seem quite useful as  such traits have been  observed for  diffusion process of passive particles inside cellular environment, for instance, see Ref. \cite{toyota2011non}.   The  method we have addressed here can be utilized, in principle, to study the dynamics in underdamped limit. In the second part we have discussed dynamics of a trapped particle.  For reasons of  simplicity and solvability,  the potential was assumed as harmonic and the particle was initially taken to be in thermal equilibrium, with no active noise present.   When the active noise is introduced into the system in the form of correlated Gaussian noise it  still remains normal distributed but with an extended standard deviation which is signature of this extra noise. Here it should be mentioned that the elevation of standard deviation can be attributed to the increase of potential energy of the system as well as the local temperature, sometimes termed as 'effective temperature' \cite{PhysRevE.77.051111,PhysRevLett.113.238303}. A confined  system in dichotomous noise replicates the dynamics of self propelled particles for which case the probability is mainly concentrated near $\pm u/\lambda$ at steady state.   In the case where the active noise is modeled as Poissonian white noise, the distribution strongly deviates from Gaussian behavior as time passes and at stationary limit it attains a distribution which decays exponentially with distance from equilibrium position. This is an obvious demonstration of out-of-equilibrium state. Clearly this system is an ideal choice to study  useful thermodynamic properties ($e.g.$, heat, work, entropy) of optically trapped particle in an active bath \cite{goswami2019heat,GOSWAMI2019223}.  

\section{Acknowledgments}
Koushik  would like to acknowledge IISc for financial support.  The work of KLS was supported by the J.C. Bose Fellowship of the Department of Science and Technology, Govt. of India.  
\appendix

\section{ Limiting forms of PDF in the case where the active noise is dichotomous
\label{appen1} }

In the following, we study the short term, as well as long term limits
of the probability distribution given as a convolution in Eq. (\ref{G'}).
The first term of inverse  Fourier transform (IFT) given in Eq. (\ref{g2'}) is a delta function.
So its convolution with Eq. (\ref{g1'}) is easy and the result is 
\begin{align}
P_1=&\sqrt{\frac{\epsilon_A }{4\pi \tilde{T}}}\int_{-\infty}^{+\infty}d\tilde{x}'\,e^{-\frac{\epsilon_A (\tilde{x}-\tilde{x}')^{2}}{4 \tilde{T}}}\,\frac{e^{-\frac{\tilde{T}}{2}}}{2}[\delta(\tilde{x}'+\tilde{ T})+\delta(\tilde{x}'-\tilde{ T})]\nonumber\\
&=\sqrt{\frac{\epsilon_A\,e^{-\tilde{T}} }{16\pi \tilde{T}}}\left[e^{-\frac{\epsilon_A (\tilde{x}-\tilde{ T})^{2}}{4\tilde{T}}}+e^{-\frac{\epsilon_A (\tilde{x}+\tilde{ T})^{2}}{4\tilde{T}}}\right]\label{p1}.
\end{align}
The second and third term of IFT in Eq. (\ref{g2'}) contain modified
Bessel functions of first kind of zero and first order, respectively.
One can write the Modified Bessel function $I_{n}(x)$ of integer
order $n$, as an integral representation \cite{bender2013advanced}
\begin{equation}
I_{n}(x)=\frac{1}{\pi}\int_{0}^{\pi}d\theta\,e^{x\, \text{cos}\theta}\,\text{cos}n\theta\label{eq:bessel}.
\end{equation}
The evaluation of the term involving $I_{0}$ is discussed below.
\begin{align*}
P_{2}= & \sqrt{\frac{\epsilon_A }{4\pi \tilde{T}}}\int_{-\infty}^{+\infty}d\tilde{x}'\,e^{-\frac{\epsilon_A (\tilde{x}-\tilde{x}')^{2}}{4 \tilde{T}}}\,\frac{e^{-\frac{\tilde{T}}{2}}}{4}[\theta(\tilde{x}'+\tilde{T})-\theta(\tilde{x}'-\tilde{T})]I_{0}\left[\frac{\tilde{ T}}{2}\sqrt{1-\left(\frac{\tilde{x}'}{\tilde{T}}\right)^{2}}\right]\\
= & \sqrt{\frac{\epsilon_A\,e^{-\tilde{T}} }{64\,\pi \tilde{T}}}\int_{-\tilde{T}}^{+\tilde{T}}d\tilde{x'}\,e^{-\frac{\epsilon_A (\tilde{x}-\tilde{x}')^{2}}{4 \tilde{T}}}\,I_{0}\left[\frac{\tilde{ T}}{2}\sqrt{1-\left(\frac{\tilde{x}'}{\tilde{T}}\right)^{2}}\right]
\end{align*}
After taking $\frac{\tilde{x}'}{\tilde{T}}=\text{sin}\phi$ and rewriting $P_{1}$ using
relation (\ref{eq:bessel}) one can arrive at
\begin{align}
  P_{2}&=\frac{1}{\pi}\sqrt{\frac{\epsilon_A \tilde{T}\,e^{-\tilde{T}} }{64\,\pi }}\int_{-\frac{\pi}{2}}^{+\frac{\pi}{2}}d\phi\,\text{cos}\phi\int_{0}^{\pi}d\theta\,e^{-\frac{\epsilon_A (\tilde{x}-\tilde{T}\text{sin}\phi)^{2}}{4 \tilde{T}}}\,e^{\frac{\tilde{ T}}{2}\text{cos}\phi\,\text{cos}\theta}\nonumber\\
 &=\frac{1}{\pi}\sqrt{\frac{\epsilon_A \tilde{T}\,e^{-\tilde{T}} }{64\,\pi }}\int_{-\frac{\pi}{2}}^{+\frac{\pi}{2}}d\phi\,\text{cos}\phi\int_{0}^{\pi}d\theta\,e^{\frac{\tilde{T}}{2}\text{cos}\phi\,\text{cos}\theta-\frac{\epsilon_A\,\tilde{x}^{2}}{4\tilde{T}}-\frac{\epsilon_A\,\tilde{T}\text{sin}^{2}\phi}{4}+\frac{\epsilon_A\,\tilde{x}}{2}\text{sin}\phi}\label{p2_1}.
\end{align}
Similarly for term involving $I_{1}(x)$,
\begin{align}
P_{3}= & \sqrt{\frac{\epsilon_A }{4\pi \tilde{T}}}\int_{-\infty}^{+\infty}d\tilde{x}'\,e^{-\frac{\epsilon_A (\tilde{x}-\tilde{x}')^{2}}{4 \tilde{T}}}\,\frac{e^{-\frac{\tilde{T}}{2}}}{4}\,\frac{[\theta(\tilde{x}'+\tilde{T})-\theta(\tilde{x}'-\tilde{T})]}{\sqrt{1-\left(\frac{\tilde{x}'}{\tilde{T}}\right)^{2}}}\,I_{1}\left[\frac{\tilde{ T}}{2}\sqrt{1-\left(\frac{\tilde{x}'}{\tilde{T}}\right)^{2}}\right]\nonumber\\
= & \sqrt{\frac{\epsilon_A\,e^{-\tilde{T}} }{64\,\pi \tilde{T}}}\int_{-\tilde{T}}^{+\tilde{T}}d\tilde{x'}\,\frac{e^{-\frac{\epsilon_A (\tilde{x}-\tilde{x}')^{2}}{4 \tilde{T}}}}{\sqrt{1-\left(\frac{\tilde{x}'}{\tilde{T}}\right)^{2}}}\,I_{1}\left[\frac{\tilde{ T}}{2}\sqrt{1-\left(\frac{\tilde{x}'}{\tilde{T}}\right)^{2}}\right]\nonumber\\
=&\frac{1}{\pi}\sqrt{\frac{\epsilon_A \tilde{T}\,e^{-\tilde{T}} }{64\,\pi }}\int_{-\frac{\pi}{2}}^{+\frac{\pi}{2}}d\phi\int_{0}^{\pi}d\theta\,\text{cos}\theta\,e^{\frac{\tilde{T}}{2}\text{cos}\phi\,\text{cos}\theta-\frac{\epsilon_A\,\tilde{x}^{2}}{4\tilde{T}}-\frac{\epsilon_A\,\tilde{T}\text{sin}^{2}\phi}{4}+\frac{\epsilon_A\,\tilde{x}}{2}\text{sin}\phi}\label{p3_1}.
\end{align}
\subsubsection{Small time limit}
Let us consider the limit $\tilde{T}\ll 1.$ At this temporal regime in Eq. (\ref{p2_1})  we can expand the term $e^{\frac{\tilde{T}}{2}\text{cos}\phi\,\text{cos}\theta}$  and keeping only first two terms one can approximate $P_2$ as follows:
\begin{align}
P_2=&\frac{1}{\pi}\sqrt{\frac{\epsilon_A \tilde{T}\,e^{-\tilde{T}} }{64\,\pi }}\int_{-\frac{\pi}{2}}^{+\frac{\pi}{2}}d\phi\,\text{cos}\phi\int_{0}^{\pi}d\theta\,\left(1+\frac{\tilde{T}}{2}\text{cos}\phi\,\text{cos}\theta\right)e^{-\frac{\epsilon_A\,\tilde{x}^{2}}{4\tilde{T}}-\frac{\epsilon_A\,\tilde{T}\text{sin}^{2}\phi}{4}+\frac{\epsilon_A\,\tilde{x}}{2}\text{sin}\phi}\nonumber\\
=& \sqrt{\frac{\epsilon_A \tilde{T}\,e^{-\tilde{T}} }{64\,\pi }}\int_{-\frac{\pi}{2}}^{+\frac{\pi}{2}}d\phi\,\text{cos}\phi\,e^{-\frac{\epsilon_A\,\tilde{x}^{2}}{4\tilde{T}}-\frac{\epsilon_A\,\tilde{T}\text{sin}^{2}\phi}{4}+\frac{\epsilon_A\,\tilde{x}}{2}\text{sin}\phi}\nonumber\\
=&\sqrt{\frac{\epsilon_A \tilde{T}\,e^{-\tilde{T}} }{64\,\pi }}e^{-\frac{\epsilon_A\,\tilde{x}^{2}}{4\tilde{T}}}\int_{-1}^{+1}dz\,e^{-\frac{\epsilon_A\,\tilde{T}z^{2}}{4}+\frac{\epsilon_A\,\tilde{x}}{2}z}\nonumber\\
=& \sqrt{\frac{e^{-\tilde{T}} }{64}}\,\left[\text{erf}\left(\sqrt{\frac{\epsilon_A}{4 \tilde{T}}}(\tilde{x}+\tilde{T})\right)-\text{erf}\left(\sqrt{\frac{\epsilon_A}{4 \tilde{T}}}(\tilde{x}-\tilde{T})\right)\right]\label{p2_2}
\end{align}
Likewise,
\begin{align}
P_3=&\frac{1}{\pi}\sqrt{\frac{\epsilon_A \tilde{T}\,e^{-\tilde{T}} }{64\,\pi }}\int_{-\frac{\pi}{2}}^{+\frac{\pi}{2}}d\phi\int_{0}^{\pi}d\theta\,\text{cos}\theta\,\left(1+\frac{\tilde{T}}{2}\text{cos}\phi\,\text{cos}\theta\right)e^{-\frac{\epsilon_A\,\tilde{x}^{2}}{4\tilde{T}}-\frac{\epsilon_A\,\tilde{T}\text{sin}^{2}\phi}{4}+\frac{\epsilon_A\,\tilde{x}}{2}\text{sin}\phi}\nonumber\\
=&\frac{\tilde{T}}{4} \sqrt{\frac{\epsilon_A \tilde{T}\,e^{-\tilde{T}} }{64\,\pi }}\int_{-\frac{\pi}{2}}^{+\frac{\pi}{2}}d\phi\,\text{cos}\phi\,e^{-\frac{\epsilon_A\,\tilde{x}^{2}}{4\tilde{T}}-\frac{\epsilon_A\,\tilde{T}\text{sin}^{2}\phi}{4}+\frac{\epsilon_A\,\tilde{x}}{2}\text{sin}\phi}\nonumber\\
=& \frac{\tilde{T}}{4}\sqrt{\frac{e^{-\tilde{T}} }{64}}\,\left[\text{erf}\left(\sqrt{\frac{\epsilon_A}{4 \tilde{T}}}(\tilde{x}+\tilde{T})\right)-\text{erf}\left(\sqrt{\frac{\epsilon_A}{4 \tilde{T}}}(\tilde{x}-\tilde{T})\right)\right]=\mathcal{O}(\tilde{T}^2)\label{p3_2}.
\end{align}

Therefore, by virtue of Eq. (\ref{p1}),(\ref{p2_2}),(\ref{p3_2}) the probability
distribution function at very short time limit  is obtained as

\begin{align}
\mathbb{P}(\tilde{x}_{f},\tilde{T};\,\tilde{x}_{0}=0,0)&=\sqrt{\frac{\epsilon_A\,e^{-\tilde{T}} }{16\pi \tilde{T}}}\left[e^{-\frac{\epsilon_A (\tilde{x}_f-\tilde{ T})^{2}}{4\tilde{T}}}+e^{-\frac{\epsilon_A (\tilde{x}_f+\tilde{ T})^{2}}{4\tilde{T}}}\right]\nonumber\\
&+\sqrt{\frac{e^{-\tilde{T}} }{64}}\,\left[\text{erf}\left(\sqrt{\frac{\epsilon_A}{4 \tilde{T}}}(\tilde{x}_f+\tilde{T})\right)-\text{erf}\left(\sqrt{\frac{\epsilon_A}{4 \tilde{T}}}(\tilde{x}_f-\tilde{T})\right)\right]\label{P dicho small T}
\end{align}
 \subsubsection{Long time limit}
For large $s$, Modified Bessel function can be expanded asymptotically as
\begin{equation}
I_{n}(s)\sim\sqrt{\frac{1}{2\pi s}}\,e^{s}\quad\text{for}\:s\rightarrow\infty\label{eq:asymp bessel}.
\end{equation}
Using Eq. (\ref{eq:asymp bessel}) we can write down $I_n(x)$ at long $\tilde{T}$ limit as
\begin{align*}
I_{n}\left[\frac{\tilde{ T}}{2}\sqrt{1-\left(\frac{\tilde{x}'}{\tilde{T}}\right)^{2}}\right]& \sim \sqrt{\frac{1}{2 \pi}}\,\frac{e^{\frac{\tilde{ T}}{2}\sqrt{1-\left(\frac{\tilde{x}'}{\tilde{T}}\right)^{2}}}}{\sqrt{\frac{\tilde{ T}}{2}\sqrt{1-\left(\frac{\tilde{x}'}{\tilde{T}}\right)^{2}}}}\\
&\sim \sqrt{\frac{1}{\pi \tilde{T}}}\,e^{\frac{\tilde{ T}}{2}\left(1-\frac{1}{2}\left(\frac{\tilde{x}'}{\tilde{T}}\right)^{2}\right)}\\
&=\sqrt{\frac{e^{\tilde{T}}}{\pi \tilde{T}}}\,e^{-\frac{\tilde{x}'^2}{4 \tilde{T}}}.
\end{align*} 
So, at the limit $\tilde{T}\gg 1$, $P_2$ becomes
\begin{align}
P_2=&\underset{\tilde{T}\rightarrow \infty}{\mathcal{L}t}\sqrt{\frac{\epsilon_A\,e^{-\tilde{T}} }{64\,\pi \tilde{T}}}\sqrt{\frac{e^{\tilde{T}}}{\pi \tilde{T}}}\int_{-\tilde{T}}^{+\tilde{T}}d\tilde{x'}\,e^{-\frac{\epsilon_A (\tilde{x}-\tilde{x}')^{2}}{4 \tilde{T}}}\,e^{-\frac{\tilde{x}'^2}{4 \tilde{T}}}\nonumber\\
=&\sqrt{\frac{\epsilon_A}{64\,\pi^2 \tilde{T}^2}}\,\int_{-\infty}^{+\infty}d\tilde{x'}\,e^{-\frac{\epsilon_A (\tilde{x}-\tilde{x}')^{2}}{4 \tilde{T}}}\,e^{-\frac{\tilde{x}'^2}{4 \tilde{T}}}\nonumber\\
=& \sqrt{\frac{\epsilon_A}{64\,\pi^2 \tilde{T}^2}} \sqrt{\frac{4 \pi \tilde{T}}{1+\epsilon_A}}\,e^{-\frac{\epsilon_A}{1+\epsilon_A}\frac{\tilde{x}^2}{4 \tilde{T}}}\nonumber\\
=&\frac{1}{2}\sqrt{\frac{\epsilon_A}{4(1+\epsilon_A)\,\pi \tilde{T}}}\,e^{-\frac{\epsilon_A}{1+\epsilon_A}\frac{\tilde{x}^2}{4 \tilde{T}}}.
\end{align}
In similar fashion,
\begin{align}
P_3=\frac{1}{2}\sqrt{\frac{\epsilon_A}{4(1+\epsilon_A)\,\pi \tilde{T}}}\,e^{-\frac{\epsilon_A}{1+\epsilon_A}\frac{\tilde{x}^2}{4 \tilde{T}}}.
\end{align}
For $\tilde{T}\rightarrow \infty$, 
\begin{align*}
P_1=\underset{\tilde{T}\rightarrow \infty}{\mathcal{L}t}\sqrt{\frac{\epsilon_A\,e^{-\tilde{T}} }{16\pi \tilde{T}}}\left[e^{-\frac{\epsilon_A (\tilde{x}-\tilde{ T})^{2}}{4\tilde{T}}}+e^{-\frac{\epsilon_A (\tilde{x}+\tilde{ T})^{2}}{4\tilde{T}}}\right]\rightarrow 0.
\end{align*}

Therefore, from Eq. (\ref{G'}) one can obtain the probability distribution function at large time limit as

\begin{equation}
\mathbb{P}(\tilde{x}_{f},\tilde{T};\,\tilde{x}_{0}=0,0)=\sqrt{\frac{\epsilon_A}{4(1+\epsilon_A)\,\pi \tilde{T}}}\,e^{-\frac{\epsilon_A}{1+\epsilon_A}\frac{\tilde{x}^2}{4 \tilde{T}}}\label{P dicho large T}.
\end{equation}

\section{ Moments of  $\sigma_{PC}(t)$ and its expression in terms of $\sigma_{PW}(t)$
\label{appen3} }
 Using Eq. (\ref{eq:active force dynamics}) one can express $\sigma_{PC}(t)$
in terms of $\sigma_{PW}(t)$ as 
\begin{align*}
\sigma_{PC}(t) & =  \lambda_p \int_{-\infty}^{t}ds\,e^{-\lambda_p (t-s)}\sigma_{PW}(s)\nonumber\\
 & =
 \lambda_p  \int_{-\infty}^{+\infty}ds\,\Theta(t-s)\,e^{-\lambda_p (t-s)}\sigma_{PW}(s).
\end{align*}
So the second order correlation can be written as
\begin{align}
\left<\sigma_{PC}(t_1)\sigma_{PC}(t_2)\right>=\lambda_p ^2\int_{-\infty}^{+\infty}ds_1\int_{-\infty}^{+\infty}ds_2\,&\Theta(t_1-s_1)\,\Theta(t_2-s_2)\,e^{-\lambda_p (t_1-s_1+t_2-s_2)}\nonumber\\
&\left<\sigma_{PW}(s_1)\sigma_{PW}(s_2)\right> \label{correaltionsigmapc}.
\end{align}
Using the auto-correlation of $\sigma_{PW}(t)$ as given in Eq. (\ref{correlationpw})  and integrating over $s_1$ we reach at
\begin{align}
\left<\sigma_{PC}(t_1)\sigma_{PC}(t_2)\right>=2D_A\,\lambda_p ^2\int_{-\infty}^{+\infty}ds_2\,\Theta(t_1-s_2)\,\Theta(t_2-s_2)\,e^{-\lambda_p (t_1+t_2-2s_2)}\label{correaltionsigmapc_1},
\end{align}
which follows
\begin{align}
\left<\sigma_{PC}(t_1)\sigma_{PC}(t_2)\right>=&2D_A\,\lambda_p ^2\int_{-\infty}^{\text{min}(t_1,t_2)}ds_2\,\Theta(t_1-s_2)\,\Theta(t_2-s_2)\,e^{-\lambda_p (t_1+t_2-2s_2)}\nonumber\\
=&D_A \,\lambda_p \, e^{-\lambda_p  \vert t_1-t_2\vert}\nonumber\\
=&\frac{D_A}{\tau_p}\, e^{-\vert t_1-t_2\vert/\tau_p}\label{correlationsigmapcfinal}.
\end{align}
 $\tau_{p}$ is correlation (or persistence) time of active noise which is defined as $\tau_{p}=\frac{1}{\lambda_p }$. Note that, at the limit $\tau_p \rightarrow 0$, $\sigma
 _{PC}$ tends to $\sigma
 _{PW}$ as it is anticipated from Eq. (\ref{eq:active force dynamics}) and the correlation becomes delta-correlated as one can obtain from Eq. (\ref{correlationsigmapcfinal}), taking the said limit.

In Eq. (\ref{eq:ch zeta(t)}) we have expressed characteristic function of $\sigma_{PC}(t)$ in terms of $\sigma_{PW}(t)$ which we have obtained as follows:
\begin{align}
\int_{0}^{T}dt\,\sigma_{PC}(t)  = & \lambda_p \int_{0}^{T}dt\int_{-\infty}^{+\infty}ds\,\Theta(t-s)\,e^{-\lambda_p (t-s)}\sigma_{PW}(s)\nonumber \\
=& \lambda_p \int_{-\infty}^{+\infty}ds\,\int_{-\infty}^{+\infty}dt\,\Theta(t-s)\Theta(T-s)\Theta(t)\Theta(T-t)\,e^{\lambda_p  s}e^{-\lambda_p  t}\,\sigma_{PW}(s)\nonumber
\\
=&\int_{-\infty}^{+\infty}ds\,\Theta(T-s)\,e^{\lambda_p  s}\,\sigma_{PW}(s)\left(e^{-\lambda_p \,\text{max}(0,s)}-e^{-\lambda_p \,T}\right)\nonumber
\\
  = & \int_{-\infty}^{+\infty}ds\,e^{\lambda_p  s}(1-e^{-\lambda_p  T})\,\sigma_{PW}(s)\:;\mbox{ if }s<0\nonumber \\
  = & \int_{-\infty}^{+\infty}ds\,(1-e^{-\lambda_p (T-s)})\,\sigma_{PW}(s)\:;\mbox{ if }T\geq s\geq0.\label{eq:zeta(t)-Sigma(t)}
\end{align}

\section{ Calculation of stationary probability distribution function for $\tilde{\tau}=0.5$ \label{appen 1}}
Stationary distribution can be obtained by considering the limit $\tilde{T}\rightarrow \infty$. In this limit, Eq. (\ref{Probfinsho_dless}) becomes
\begin{align}
\mathbb{P}(\tilde{x}_{f})=\frac{1}{2\pi}\int_{-\infty}^{+\infty}d\tilde{p}_{T}\,e^{-i\tilde{p}_{T}\,\tilde{x}_{f}}\,e^{-\frac{\tilde{p}_{T}^{2}}{2\epsilon_A\,\tilde{\tau}}}\,\left[\frac{1}{1+\tilde{p}_{T}^{2}}\right]^{\frac{1}{2\tilde{\tau}}}.
\end{align}
Using the identity 
\begin{equation}
\left[\frac{1}{1+\tilde{p}_{T}^{2}}\right]^{\frac{1}{2\tilde{\tau}}}=\frac{1}{\Gamma(\frac{1}{2\tilde{\tau}})}\int_{0}^{\infty}e^{-(1+\tilde{p}_{T}^{2})s}s^{\frac{1}{2\tilde{\tau}}-1}ds\label{relation gamma},
\end{equation}
 in the above equation and doing the Fourier inversion with respect to $\tilde{p}_{T}$ (which is essentially a Gaussian integration) we arrive at
\begin{equation}
\mathbb{P}(\tilde{x}_{f})=\frac{1}{\Gamma(\frac{1}{2\tilde{\tau}})}\int_{0}^{\infty}\frac{s^{\frac{1}{2\tilde{\tau}}-1}\,e^{-s-\frac{\tilde{x}_{f}^{2}}{4(\alpha+s)}}}{\sqrt{4\pi(\alpha+s)}}ds\label{eq:P s-integrand},
\end{equation}
where $\alpha=\frac{1}{2\epsilon_A\,\tilde{\tau}}$ .
For $2\tilde{\tau}=1$, the above integration can be performed exactly which
leads to the following result:
\begin{equation}
\mathbb{P}(\tilde{x}_{f})=\frac{1}{4}e^{-\mid\tilde{ x}_{f}\mid+\alpha}\left[2-\text{erfc}\left(\frac{\mid \tilde{x}_{f}\mid-2\alpha}{2\sqrt{\alpha}}\right)+e^{2\mid \tilde{x}_{f}\mid}\,\text{erfc}\left(\frac{\mid \tilde{x}_{f}\mid+2\alpha}{2\sqrt{\alpha}}\right)\right]\label{P for Nu 1 }.
\end{equation}


\end{document}